 \documentclass[twocolumn,showpacs]{revtex4}

\usepackage[latin1]{inputenc}
\usepackage{graphicx}%
\usepackage{dcolumn}
\usepackage{amsmath}

\makeatletter
\def\btt#1{\texttt{\@backslashchar#1}}%
\DeclareRobustCommand\bblash{\btt{\@backslashchar}}%
\makeatother

\begin{document}

  \title{A Methodology of Cooperative Driving  based on Microscopic Traffic Prediction}

\author{Boris S. Kerner$^1$, Sergey L. Klenov$^2$, Vincent Wiering$^1$,  and Michael Schreckenberg$^1$}
 
  \affiliation{$^1$
Physics of Transport and Traffic, University of Duisburg-Essen,
47048 Duisburg, Germany}

   \affiliation{$^2$
Moscow Institute of Physics and Technology,
Department of Physics, 141700 Dolgoprudny, Moscow Region, Russia}


 \begin{abstract}
We present a methodology of cooperative driving   in  vehicular traffic,  in which for short-time traffic prediction
rather than one of the  statistical approaches of artificial intelligence (AI),    
 we follow a qualitative different microscopic traffic prediction approach    developed recently
[Phys. Rev. E  106  (2022) 044307].  In the microscopic traffic prediction approach used for the planning of the subject  vehicle trajectory,   no learning algorithms of  AI  are applied; instead, microscopic traffic modeling    based on   the physics of vehicle motion is used.  The presented methodology of cooperative driving 
is devoted to application cases in which microscopic traffic prediction   without   cooperative driving
cannot lead to a successful   vehicle control and trajectory planning.  For the understanding of the physical features of the methodology of cooperative driving,
a   traffic city scenario has been numerically studied, 
 in which a subject vehicle, which requires cooperative driving, is an automated vehicle. 
  Based on    microscopic traffic prediction, in the methodology first a cooperating vehicle(s) is found; 
 then, motion requirements for the cooperating vehicle(s)   and
 characteristics of automated vehicle control are predicted and used for vehicle motion;
to update   predicted characteristics of vehicle motion,   calculations of  the
predictions of motion requirements for the cooperating vehicle   and
 automated vehicle control are repeated for each next time instant at which
new  measured data for current microscopic traffic situation are available.
 With the use of microscopic traffic simulations,  the evaluation of the applicability  
of this methodology is illustrated
for a simple case of  unsignalized city intersection,
 when the automated   vehicle  wants to turn right from
a secondary road onto the priority road.  
\end{abstract}

\maketitle

\section{Introduction  \label{Int}}

\subsection{Brief Overview of Approaches to Cooperative Driving   }	

Cooperative driving    is currently one of the most developing fields in physics of vehicular traffic.
For the realization of cooperative driving,  a vehicular networking is usually needed,
i.e., a vehicular network in which vehicles can communicate to each other   through vehicle-to-vehicle (V2V) communication or/and through vehicle-to-infrastructure communication; both types of vehicular communications 
are called  V2X-communication (see, e.g.,~\cite{Schmitz2004,Maurer2006,Chen2006,Wang2020,Chen2015}). 
Through V2X-communication   vehicles can have additional information about the behaviors of other vehicles. 

In particular, it is assumed that  cooperative driving should support the motion of 
 automated-driving vehicles (automated vehicles for short)   (see, e.g.,~\cite{Ioannou1997,Ioannou1996,Ioannou1993,Meyer2014,Liang2000,Davis2014,Xin_Chang2020A,Jiang-Wang2021A,Coop_AV_Papageorgiou-Mountakis_2021A,Yao-Gu2022A,Jiang-Sun2023A,Ma-Wang2021A,Luo-Gu2022A,Ma-Qu2023A,Monteiro-Ioannou2023A,Weijie-Yu2023A,Li-Xiao2023A,Gu-Wang2023A,Coop_AV_P-Hang_2023A,Lane-Changing_AV_2024A,Coop_AV_J-Shi_2023A,Coop_Cont_AV_2024A,Coop_AV_K-Hou_2023A,Coop_Cont_AV_Huang_2024B,Coop_AV_Ying-Shang_2024A,AV_KKSW_Liu_2023A,Coop_AV_Jiahua-Qiu_2023A,Mixed_Maiti_2023A,Coop_AV_L-An_2023A,Coop_AV_Zhang_2023A,Mixed_Wang_2023A,Mixed_Lihua-Luo_2024A,Mixed_Lin-Hou_2023A,Mixed_Ziyu-Cui_2023A,Coop_AV_Futao-Zhang_2023A,Coop_AV_Shi-Teng-Zheng_2023A,Coop_AV_Leyi-Duan_2023A,Coop_AV_Yangsheng-Jiang_2023A,Coop_AV_Ruijie-Li_2023A,Coop_AV_C-Liu_2023A,Coop_AV_Dian-Jing_2023A,Coop_AV_X-Huang_2023A,Lane-Changing_AV_Xia-Li_2023A,Coop_AV_Jiawei-Lu_2023A,Coop_AV_Yu-Du_2023A,Coop_AV_Yunjie-Liu_2023A,Coop_AV_I-Gokasar_2023A,Coop_AV_Jiawei-Wang_2023A,Coop_AV_Jie-Wang_2023A,Coop_AV_Z-Huang_2023A,Coop_AV_Mengting-Guo_2023A,Coop_AV_Qishen-Zhou_2023A,Coop_AV_Yan-Tao-Zhang_2023A,Coop_AV_W-Yue_2023A,Coop_AV_Chenguang-Zhao_2023A,Coop_AV_Minghao-Fu_2023A,Coop_AV_Cong-Zhai_2023A,Coop_AV_Yanyan-Qin_2023A,Lane-Ch_X-Duan-C_2023A,Z-Huang-H_2023A,Coop_AV_S-S-Mousavi_2023A,Coop_AV_Weijie-Yu_2023A,Coop_AV_J-Wen-S_2023A,Coop_AV_Matin-Dia_2023A,Coop_AV_Milad-Malekzadeh_2024A,Coop_AV_Ying-Feng_2024A}).  
 The information obtained through V2X-communication
   can also be used for a short-time prediction of a traffic situation, for example, based on a so-called partially observable Markov decision process (see, e.g.,~\cite{Brechtel2014,Lin2019,Schorner2019,Klimenko2014,Hubmann2019,A-Lombard-A_2023A,J-Luo-T_2023A,Markov_Shi_2024A}), learning algorithms (see, e.g.,~\cite{Isele2018,Qiao2018,Sama2020,D-Chen_2023A,X-Jiang-J_2023A,Learn_Pred_AV_2024A,Zhou-Z-Cao_2023A,Zhang-S-Li_2023A,Shupei-Wang_2024A,Renteng-Yuan_2023A,Du-Y-Zou_2023A,K-Guo-M-Wu_2023A,Chunyu-Liu_2023A,J-Guo_2023A,Bharti-Redhu_2023A,Wensong-Zhang_2023A,Shun-Wang_2023A,Siyuan-Feng_2023A,Li-Z-Huang_2023A,Jiaxin-Liu_2023A,Zelin-Wang_2024A,C-J-Hoel_2023A,S-Fang-C_2023A}) as well as many other artificial intelligence (AI) and
model approaches (see, e.g.,~\cite{Orzechowski2018,Althoff2016,Naumann2019,Tas2018,Akagi2015,Morales2017,Yoshihara2017,Takeuchi2015,Yu2019,Hoermann2017,Guojing-Hu_2023A,Changxi-Ma_2023A,S-Akhtar_2023A,Xiaoxue-Yang_2023A,Gao-X-Li_2023A,Z-Gu_2023A,Y-Zhang_2023A,Xiaoyong-Sun_2023A,Weibin-Zhang_2023A,H-Chen-Y-Liu_2023A,W-Shao_2023A,T-Qie-W_2023A}). The short-time prediction
 of vehicle variables (vehicle locations and speeds)    is required for the vehicle  trajectory planning.

Thus,  the short-time prediction of  vehicle variables (vehicle locations and speeds), which is required for the   vehicle  trajectory planning, is usually made based on a diverse variety of {\it statistical} approaches of AI. Some of these statistical approaches  have been mentioned above~\cite{Brechtel2014,Lin2019,Schorner2019,Klimenko2014,Hubmann2019,A-Lombard-A_2023A,J-Luo-T_2023A,Markov_Shi_2024A,Isele2018,Qiao2018,Sama2020,D-Chen_2023A,X-Jiang-J_2023A,Learn_Pred_AV_2024A,Zhou-Z-Cao_2023A,Zhang-S-Li_2023A,Shupei-Wang_2024A,Renteng-Yuan_2023A,Du-Y-Zou_2023A,K-Guo-M-Wu_2023A,Chunyu-Liu_2023A,J-Guo_2023A,Bharti-Redhu_2023A,Wensong-Zhang_2023A,Shun-Wang_2023A,Siyuan-Feng_2023A,Li-Z-Huang_2023A,Jiaxin-Liu_2023A,Zelin-Wang_2024A,C-J-Hoel_2023A,S-Fang-C_2023A,Orzechowski2018,Althoff2016,Naumann2019,Tas2018,Akagi2015,Morales2017,Yoshihara2017,Takeuchi2015,Yu2019,Hoermann2017,Guojing-Hu_2023A,Changxi-Ma_2023A,S-Akhtar_2023A,Xiaoxue-Yang_2023A,Gao-X-Li_2023A,Z-Gu_2023A,Y-Zhang_2023A,Xiaoyong-Sun_2023A,Weibin-Zhang_2023A,H-Chen-Y-Liu_2023A,W-Shao_2023A,T-Qie-W_2023A}.

For reasons that will be explained in Sec.~\ref{MatMod_vs_AI}, rather than one of the  statistical approaches of AI, in this paper
we follow a qualitative different approach, which is based on the physics of vehicle motion in vehicular traffic. In this approach introduced in~\cite{KKl2022_Pred}, {\it no} statistical analysis of a historical  traffic database  is used for traffic prediction. 
Instead of statistical analysis of  empirical traffic data, microscopic traffic modeling
with a traffic flow model that can reproduce spatiotemporal features of real traffic is applied.
Based on   this microscopic traffic modeling,
the microscopic short-time prediction of   vehicle variables (vehicle locations and speeds) is made. This prediction is used for 
   the planning of the  vehicle trajectory.
Because  the microscopic traffic prediction of~\cite{KKl2022_Pred} is the scientific basis of the
methodology of cooperative driving presented in this paper, for the paper understanding we present a brief summary of this microscopic 
traffic prediction approach below.

\subsection{Brief Summary of Microscopic Traffic Prediction }
 \label{Brief_MP_Sec}

General features of the  microscopic traffic prediction approach
of~\cite{KKl2022_Pred} are as follows:
\begin{enumerate}
\item [(i)] We assume  that through vehicular networking (in particular, V2X-communication) at some time instants 
\begin{equation}
t=t_{p},\ p=1,2,3,\ldots
\label{t_p}
\end{equation}
  locations (including the correspondence to road lanes) and speeds of all vehicles moving around of a subject vehicle are known.   We call the multitude of the vehicle locations  $x$  and   vehicle  speeds $v$
		as {\it a microscopic traffic situation} at   time instant $t_{p}$.  In general, a time interval $\delta t=t_{p+1}-\ t_{p},\ p=1,2,3,\ldots$ that can be very short (e.g., between 0.1 s and 1 s) should not be necessarily a constant value; this time interval is determined by the measurement technology of the microscopic traffic situation. However, for simulations made in~\cite{KKl2022_Pred} as well as   in this paper
		we apply a constant time interval 
		\begin{equation}
\delta t = t_{p+1}-t_{p}=\tau, \ p=1,2,3,\ldots
\label{time-interval_f}
\end{equation}
 where $\tau$     is a time-independent time step of the model used for microscopic simulations (in all simulations
we use $\tau =$ 1 s). 	
	\item [(ii)]	A microscopic traffic situation measured at  time instant $t=t_{p}$, i.e., values for   locations and speeds of vehicles related to  $t=t_{p}$ are set into the prediction model.  These locations and speeds of vehicles
		are used as an initial condition for a microscopic traffic flow model of  vehicular traffic.  
For this time instant $t=t_{p}$, the microscopic traffic flow model calculates future (predicted) microscopic traffic situation that the model predicts for   prediction horizon $\Delta T_{p}$.
\item [(iii)] Based on this microscopic prediction, parameters of  control  of the subject  vehicle applied
at time instant $t=t_{p}$ are calculated. The automated vehicle control application should ensure a safety planning of the subject  vehicle trajectory
based on   calculated control parameters.  
\item [(iv)]
It should be emphasized that the real application of the  calculated control parameters for the safety planning of the subject vehicle trajectory at time instant $t=t_{p}$  (\ref{t_p}) is only possible if a microscopic traffic model used for
the microscopic prediction during the prediction horizon $\Delta T_{p}, \ p=1,2,3,\ldots$ can make the related calculation during a negligible short time interval in comparison with time interval (\ref{time-interval_f}).   This is because only in this case
the microscopic traffic prediction remains to be a current one for the  calculations of subject  vehicle control used for the safety planning of the subject  vehicle trajectory at time instant $t=t_{p}$. Note that
the time-discrete traffic flow 
model     for  traffic  of~\cite{KKl2022_Pred}  used in the paper (see Sec.~\ref{Model_subsec})    
   is able to perform such calculations   
  during a negligible short time interval  
	in comparison with time interval (\ref{time-interval_f}). 
\item [(v)]
The subject  vehicle moves in accordance with calculated control parameters only during the time interval
\begin{equation}
t_{p}\leq t< t_{p+1},\ p=1,2,3,\ldots. 
\label{b_p_t_p}
\end{equation}
Indeed in accordance with item (i),  at $t=t_{p+1}$ a new measured microscopic traffic situation is available. Therefore,
both the microscopic traffic prediction and resulting parameters of subject  vehicle control
made for time instant $t=t_{p}$ become obsolete. 
\item [(vi)]
For this reason, at $t=t_{p+1}$ the measured microscopic traffic situation is used as a new initial condition for microscopic traffic prediction during prediction horizon $\Delta T_{p+1}$. The prediction is used for the calculation of new parameters of subject traffic control used at $t=t_{p+1}$. The parameters of subject  traffic control are applied only 
during time interval $t_{p+1}\leq t< t_{p+2}$. Then, at  $t=t_{p+2}$ a new measured microscopic traffic situation is used as a new initial condition for microscopic traffic prediction during prediction horizon $\Delta T_{p+2}$; the microscopic prediction   is used for the calculation of new parameters of subject  traffic control used at $t=t_{p+2}$, which are applied only during time interval $t_{p+2}\leq t< t_{p+3}$,
and so on.
\item [(vii)]
Thus in the microscopic prediction approach, both the microscopic traffic prediction and the resulting calculation
of parameters for subject  vehicle control are  repeated for each new time instant (\ref{t_p}),
whereas the control parameters are used only during the related time interval (\ref{b_p_t_p}) before a new measured microscopic traffic situation is available.
\item [(viii)]
 The  
microscopic traffic prediction approach introduced in~\cite{KKl2022_Pred} (see footnote [84] of~\cite{KKl2022_Pred})  is a general approach because it is applicable for both the control of
automated-driving and human-driving vehicles. For this reason,   the vehicle for which the
microscopic traffic prediction approach is applied has been called   above as
the  subject vehicle.   
In~\cite{KKl2022_Pred}, we have illustrated  this   general
 microscopic traffic prediction approach for the trajectory planning of 
the subject vehicle that is an automated vehicle.   Indeed, 
the trajectory planning of automated vehicles is currently one of the priority aims in traffic science. 
\end{enumerate}

\begin{figure}%[b]
\begin{center}
\includegraphics[width = 8 cm]{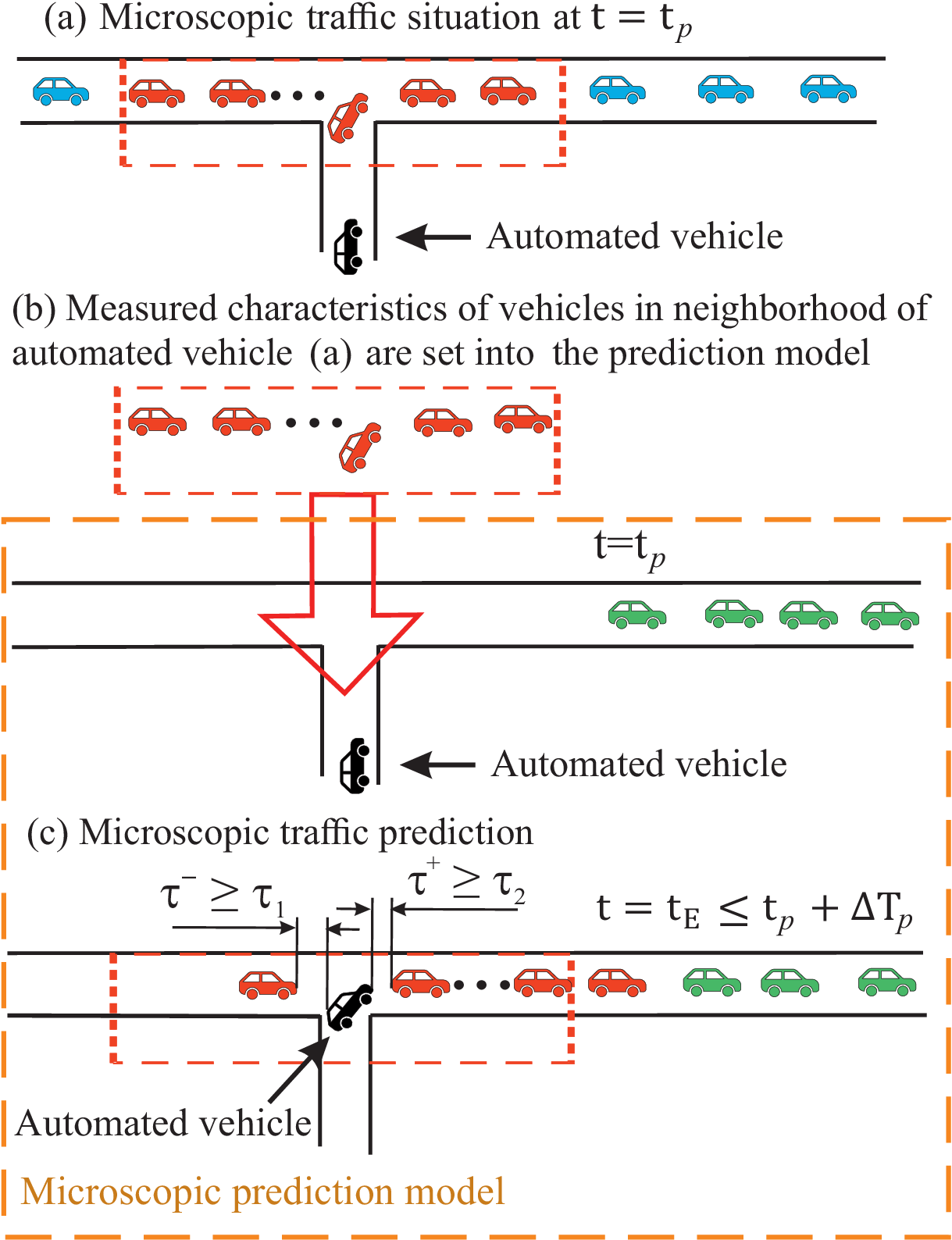} 
\end{center}
\caption{Qualitative schema of  microscopic traffic prediction  for an unsignalized intersection: (a) Microscopic traffic situation measured at $t=t_{p}$ (\ref{t_p}); only red colored vehicles that are in a red colored dashed road region can affect on the motion of the automated  vehicle (black colored vehicle).  (b) 
The measured characteristics  of red colored vehicles (vehicle speeds, vehicle locations, vehicle lengths, and so on)
 in the red colored dashed road region shown in (a) are  used
   in the microscopic prediction model; before, all vehicles that were at $t<t_{p}$  in this dashed road region of 
the microscopic prediction model are removed. (c) Through the use of the microscopic traffic prediction,  
a   predicted time instant $t_{\rm E}$ is found at which the vehicle should turn right onto the priority road. 
Adapted from~\cite{KKl2022_Pred}.}
\label{Scenario_crossing} 
\end{figure}

With the use of Fig.~\ref{Scenario_crossing}, we make a brief explanation of one of the applications of
the general features of microscopic traffic prediction approach (items (i)--(viii) above). We assume that an automated vehicle moving on a secondary road wants to turn right on the priority road at a unsignalized city intersection (Fig.~\ref{Scenario_crossing}(a)). The basic objective of the microscopic traffic prediction approach is
to   predict time dependencies of locations and speeds of vehicles in a vicinity of the interaction that
allow us to control the motion of automated vehicle. The objective of this automated vehicle control is
to make possible the turning right of the automated vehicle without the stopping at the intersection.
In this case, the parameter of automated vehicle control for the planning of automated vehicle trajectory
is deceleration/acceleration of the automated vehicle while the vehicle approaches the intersection.

	For the application scenario under consideration (Fig.~\ref{Scenario_crossing}), the calculation of the microscopic traffic prediction begins
	at some  $t=t_{p}=t_{1}$ (i.e., at $p=$ 1) at which
	a vehicle that is the preceding vehicle for the automated vehicle has just left the secondary road while turning right on the priority road. In this case, at $t=t_{1}$ control of automated vehicle can be made based on prediction of locations and speeds of vehicles moving only on the priority road
	(this case is shown in Fig.~\ref{Scenario_crossing}(a)).

	For each of the given time instant $t_{p}, \ p=1,2,3,\ldots$, in addition to the microscopic traffic prediction, the prediction model calculates
	two   predicted time instants: $t=t_{\rm min}$ and  $t=t_{\rm max}$.
	The physical sense of these predicted time instants is as follows:
	$t_{\rm min}$ and $t_{\rm max}$ are 
	predicted time instants at which the automated vehicle can reach the intersection
	moving, respectively,  with maximum acceleration (or/and maximum speed) and  
  with   a safe speed   coming to a stop
	  at $t=t_{\rm max}$.

		With the use of $t=t_{\rm min}$ and  $t=t_{\rm max}$, the prediction model searches whether there is
	a pair of vehicles moving on the priority road between which the automated vehicle can merge without stopping at the intersection while satisfying some  given safety conditions. In this case   
	\begin{equation}
	t_{\rm min} \leq t_{\rm E}< t_{\rm max},   
	\label{t_min_t_max_t_E}
	\end{equation}
	where   $t=t_{\rm E}$  is a predicted time instant at which the automated vehicle can safety turn right at the intersection
	between the above-mentioned pair of vehicles moving on the priority road;
$t_{\rm min}, t_{\rm E}, t_{\rm max} > t_{\rm 1}$.

In microscopic traffic prediction 
	made at each time instant $t_{p}$, the safety conditions at which  a pair of vehicles moving on the priority road between which the automated vehicle can merge without a stop at the intersection 
	are as follows:
	\begin{equation}
	\tau^{-}\geq \tau_{1},   \   \tau^{+}\geq \tau_{2}, 
	\label{safety_cond_basic}
	\end{equation}
	 in which
		\begin{equation}
		\tau^{-}=(s^{-}-d)/v^{-}, \ \ \tau^{+}=(s^{+}-d)/v^{\rm (AV)},
		\label{tau_-tau_+}
		\end{equation}
		where    $s^{+}$
		is a gross space gap between the automated vehicle and  the preceding vehicle;
	  $s^{-}$ is a gross space gap between the automated vehicle  
		  and the following vehicle;
		$v^{-}$ is the speed of the following vehicle; $v^{\rm (AV)}$ is the speed of the automated vehicle; all values in (\ref{tau_-tau_+}) are related to 
		a predicted time instant $t=t_{\rm E}$  at which the automated vehicle could turn right at the intersection.
		In (\ref{safety_cond_basic}),   $\tau^{-}$   is
  time headway  of a vehicle following the automated vehicle on the priority road just after the merging,  $\tau^{+}$ is time headway of the automated vehicle to the preceding vehicle on the priority road just after the merging, $\tau_{1}$ and $\tau_{2}$ are model parameters. The predicted time instant $t=t_{\rm E}$, at which the automated vehicle can merge without a stop at the intersection, 
 in addition  to safety conditions (\ref{safety_cond_basic}),    should satisfy  conditions (\ref{t_min_t_max_t_E}).
	
  Such a case, when safety conditions (\ref{safety_cond_basic})
		are satisfied, is shown in Fig.~\ref{Scenario_crossing}(c). Only in this case, predicted time instant  $t=t_{\rm E}$  is used for the calculation
		of automated vehicle deceleration. Otherwise, 
	  when at least one of the  safety conditions (\ref{safety_cond_basic}) is not satisfied, the automated vehicle stops
at the intersection. 

		 The  predicted time instant  $t=t_{\rm E}$  is used for the calculation
		of automated vehicle deceleration that is applied at $t=t_{\rm 1}$. In accordance with item (v)    of the general
		approach to microscopic traffic prediction, the automated vehicle moves in accordance with calculated deceleration only during the time interval
$t_{1}\leq t< t_{2}$. At time instant $t=t_{2}$, a new measured microscopic traffic situation is available. Therefore,
a new microscopic traffic prediction is calculated at which new predicted time instants $t_{\rm min}$, $t_{\rm max}$, and $t_{\rm E}$
as well as a new vehicle deceleration are found. The automated vehicle moves in accordance with deceleration calculated at time instant $t=t_{2}$ only during the time interval
$t_{2}\leq t< t_{3}$. At time instant $t=t_{3}$, a new measured microscopic traffic situation is available, new predicted time instants $t_{\rm min}$, $t_{\rm max}$, and $t_{\rm E}$
as well as a new vehicle deceleration are found, and so on. After the automated vehicle has turned right,
the automated vehicle moves on the priority road in accordance with given rules of automated vehicle motion.

\subsection{About Mathematical Model used for Simulations of Cooperative Driving  
\label{Model_subsec}  }

  Mathematical formulations of~\cite{KKl2022_Pred} that are used for simulations
of cooperative driving	in the paper are as follows:  We use the same microscopic model
of  mixed traffic flow as that considered in Secs.~II C   of~\cite{KKl2022_Pred}. Model for simulations of
microscopic traffic situation has been considered in Secs.~II D of~\cite{KKl2022_Pred}. Model of motion of human-driving vehicles
has been given in   Appendix~A  of~\cite{KKl2022_Pred}.
Models of the vehicle merging onto the priority road for
human-driving vehicles and for automated vehicles (when the
microscopic traffic prediction is not used) have been given 
in Appendix~B  of~\cite{KKl2022_Pred}. Model for simulations of microscopic traffic prediction
has been taken from Sec.~III and Appendix~C   of~\cite{KKl2022_Pred}.

 It should be noted that there are a huge number of other different microscopic traffic flow models for mixed
traffic (see, e.g.,~\cite{Dharba1999,Bose2003,Kesting2007,Kesting2008,Kesting2010,Kukuchi2003,Li2002,Marsden2001,Martinez2007,Shladover1995,Shladover2012,Talebpour2016,Kerner2021,Kerner2021B})\footnote{It should be noted that in many publications
 devoted to a study of  mixed traffic  the analysis of the effect of automated driving on vehicular traffic is made
(see, e.g.,~\cite{Dharba1999,Bose2003,Kesting2007,Kesting2008,Kesting2010,Kukuchi2003,Li2002,Marsden2001,Martinez2007,Shladover1995,Shladover2012,Talebpour2016,Kerner2021,Kerner2021B}). Contrary to such publications, in our paper {\it no} effect of automated vehicles on vehicular traffic is discussed. Instead,
we apply the model of mixed traffic of Refs.~\cite{KKl2022_Pred,Kerner2021,Kerner2021B}
 for simulations of cooperative driving based on the microscopic traffic prediction. We should also
 emphasize that in Refs.~\cite{KKl2022_Pred,Kerner2021,Kerner2021B}, in which the the model of mixed traffic used in the paper has been developed, the influence of automated vehicles and human-driving vehicle on mixed traffic
 as well as   characteristics of automated vehicles have already been considered in details.}. A basic requirement to a model for simulations of
the microscopic traffic prediction is that the model of human-driving vehicles should simulate microscopic
behavior of human drivers in different traffic situations as close as possible
to real measured traffic data. For motion of human-driving vehicles in mixed traffic,
we use a version of
 a microscopic stochastic time-discrete  model of Kerner and Klenov in the framework of  
  three-phase traffic theory (see Appendix~A in~\cite{KernerBook3}); in the model, in addition to hypotheses of the three-phase traffic theory, well-known
 slow-to-start rule~\cite{TT1993,Schadschneider}, and Gipps' formulation for safe speed~\cite{Gipps,Kra_PhD,Kra} have been used;
contrary to Appendix~A in~\cite{KernerBook3},  the successive number of time steps of the
delay in acceleration of a human-driving vehicle has been  limited as made in~\cite{KKl2022_Pred} (see 
formulas (A17) and (A18) of Appendix~A of~\cite{KKl2022_Pred}).  The use of this model for human-driving vehicles
is explained as follows:
 as proven in~\cite{KernerBook3}, the  Kerner-Klenov model satisfies the above-mentioned basic requirement following from empirical features of real traffic.
 In the  traffic model, update rules of  
motion of automated driving vehicles are simulated with  a time-discrete version 
of the classical
ACC model (e.g.,~\cite{Ioannou1997,Ioannou1996,Ioannou1993,Meyer2014,Liang2000,Davis2014}) that is equivalent to Helly's  
 model~\cite{Helly_1959}. 

  The time-discrete traffic flow 
model  of~\cite{KKl2022_Pred} used in this paper for simulations 
  calculates a new microscopic traffic prediction (see items (ii) and (vi) of Sec.~\ref{Brief_MP_Sec}) for the motion of about 30 vehicles  
  during a time interval $\theta <$ 0.005 s,
which is negligible short  
in comparison with time interval $\delta t=t_{p+1}- t_{p}=\tau=$ 1 s (\ref{time-interval_f}).

\subsection{Objective,  Focus, and Organization    }

{\it The objective} of this paper is the presentation   of a methodology for cooperative driving
    vehicular traffic based on short-time microscopic traffic prediction of~\cite{KKl2022_Pred}.
It must be emphasized that contrary to the analysis of applications of microscopic traffic prediction made in~\cite{KKl2022_Pred}, the  methodology of cooperative driving  presented in this paper
is devoted to application cases in which microscopic traffic prediction   without  the use of  cooperative driving
  {\it cannot}  lead to a successful   vehicle control and trajectory planning. In other words,
the objective of the paper under consideration
 makes sense only then, if subject vehicle motion   control based on the microscopic traffic prediction
without cooperative driving {\it does not} lead to    the merging of the  subject  vehicle onto the priority road   without stopping  at the intersection. 
 For this reason,   the methodology of cooperative driving based on the microscopic traffic prediction   (Sec.~\ref{Gen}), the physics and the procedure of calculations of cooperative driving
(Sec.~\ref{Conditions}) and results of simulations of this cooperative driving (Sec.~\ref{Sim_results})
are qualitatively different from results of~\cite{KKl2022_Pred} in which no  cooperative driving has been assumed. 
We limit our 
  analysis of this methodology of cooperative driving  based on microscopic traffic prediction   by a case vehicular traffic in which
  100$\%$ of vehicles can  participate and are willing to participate in  cooperative driving.

	 The main {\it focus} of this paper is the presentation of a general methodology for cooperative driving
	based on microscopic traffic prediction. To study the physical features of this general methodology for cooperative driving, we limit simulations
	by the same simple case of unsignalized city intersection discussed in~\cite{KKl2022_Pred}
	(Fig.~\ref{Scenario_crossing}) in which
	a subject vehicle is the automated vehicle that tries to turn right from the secondary road onto the priority road without stopping at the intersection.
	Contrary to the case shown qualitatively in Fig.~\ref{Scenario_crossing} and explained above in Sec.~\ref{Brief_MP_Sec},
	we simulate a traffic scenario at which microscopic traffic prediction cannot be used for the safety vehicle merging
	without the vehicle stopping at the intersection.

The paper is organized as follows:
The general methodology of cooperative driving based on the microscopic traffic prediction   is presented in Sec.~\ref{Gen}.
A simple scenario of cooperative driving for simulations of the general methodology is discussed in Sec.~\ref{Sec_Scenario}.
 The physics   of calculations of cooperative driving has been studied in  
Sec.~\ref{Conditions}. Results of simulations of this cooperative driving are presented in Sec.~\ref{Sim_results}.
In discussion section (Sec.~\ref{Disc}), we consider limitations of cooperative driving (Sec.~\ref{Sec_Limitation}),
briefly compare applications of leaning algorithms for traffic prediction versus microscopic traffic prediction
based on the physics of vehicle motion used in the paper (Sec.~\ref{MatMod_vs_AI}), discuss other possible applications
of the general methodology of cooperative driving (Sec.~\ref{Other_Sec}),
consider the effect of data uncertainty on prediction reliability
 (Sec.~\ref{real_Sec}) as well as formulate conclusions of the paper 
 (Sec.~\ref{Conc}).

\section{Methodology of Physical Modeling of Cooperative Driving  in Vehicular Traffic  \label{Gen}}
 
 The  general methodology for cooperative driving presented here can be applied for both  
automated-driving and human-driving vehicles.
In this case, as mentioned above (see item (viii) of Sec.~\ref{Brief_MP_Sec}), a vehicle, which uses the cooperative driving, can be called as
a subject vehicle, to distinguish the vehicle from cooperating vehicles.
However, the trajectory planning of automated vehicles is one of the priority aims in traffic science. For this reason, in
the paper  we have limited the study of the physics  of
  the general methodology of cooperative driving 
	for a case in which the subject vehicle is the automated vehicle.  

The general methodology  of physical modeling of cooperative driving is as follows.
		\begin{itemize}
	\item [(i)]
	 First, microscopic traffic prediction~\cite{KKl2022_Pred} for the  subject vehicle control without the application of cooperative driving
is made, as briefly explained in Sec.~\ref{Brief_MP_Sec}. 
   
 Contrary to~\cite{KKl2022_Pred}, in this paper we consider microscopic traffic situations for which already at 
\begin{equation}
t_{p}=t_{1} 
\label{t_p_1}
\end{equation}
the calculated  microscopic traffic prediction shows that
 no   effective
 subject vehicle control is possible. Therefore,  to organize the effective
 subject vehicle control,
a cooperative driving based on  microscopic traffic prediction is applied for  subject
vehicle control.
At $t_{p}=t_{1}$ 
(\ref{t_p_1}), the following stages (ii) and (iii) of cooperative driving based on microscopic traffic prediction
are applied.

\item [(ii)] Based on the calculated  microscopic traffic prediction
considered in Sec.~\ref{Brief_MP_Sec}, between vehicles that
  are willing to cooperate  vehicles that
 are suitable   ones for the realization of cooperative driving are found. These vehicle are 
called below as   $\lq\lq$cooperating vehicles".

\item [(iii)] Microscopic traffic prediction found at $t_{p}=t_{1}$ (Sec.~\ref{Brief_MP_Sec})
  is further used to find
both motion requirements for the cooperating vehicles  and
related characteristics of  subject vehicle control.   The prediction of the vehicle locations    and vehicle  speeds   are used for the calculation of the future trajectory of the  subject   vehicle as well as all other vehicles involved in   cooperative driving.
    
The microscopic traffic prediction together with the determination of the cooperating vehicles
as well as the associated calculation of
motion requirements for the cooperating vehicles   and
 characteristics of  subject vehicle control are only possible when the microscopic model is able to calculate all these
characteristics  during negligible short time interval in comparison with time interval $\delta t=t_{p+1} - t_{p}$.
 It should be emphasized that as explained in Sec.~\ref{Brief_MP_Sec} the time-discrete traffic flow 
model     for mixed traffic  of~\cite{KKl2022_Pred}  used in this paper    
   is able to perform these calculations   
  during a negligible short time interval.

\item [(iv)] Stages (ii) and (iii) explained above are repeated at the next time instant $t_{p+1},\ p=1,2,3,\ldots$, when a next microscopic traffic situation is known. The new microscopic traffic situation is used as an initial condition for the microscopic traffic flow model. Then, the model makes the microscopic traffic prediction for vehicle locations  and vehicle speeds    during prediction horizon $\Delta T_{p+1}$ (prediction horizon $\Delta T_{p}$ can be different for different time instants  $t_{p}$).    
Stages (ii) and (iii) are iterated for each new time instants, and so on, once data for new microscopic traffic situations are available.

Thus, as   in~\cite{KKl2022_Pred} (Sec.~\ref{Brief_MP_Sec}), in the methodology for cooperative driving
 there is the repetition of the microscopic traffic prediction
  together with
the calculation of motion requirements for the cooperating vehicle
for any current time instant  $t_{p}$ (\ref{t_p}) at which microscopic traffic situations are available. This allows us to update the calculations of the trajectories of both the cooperating vehicles and
the subject vehicle during the vehicle motion  in  traffic flow.
\end{itemize}

\section{Application Scenario  of Cooperative Driving 
for Unsignalized Intersection in City Traffic   \label{Sec_Scenario}}

\begin{figure}
\begin{center}
\includegraphics[width = 8 cm]{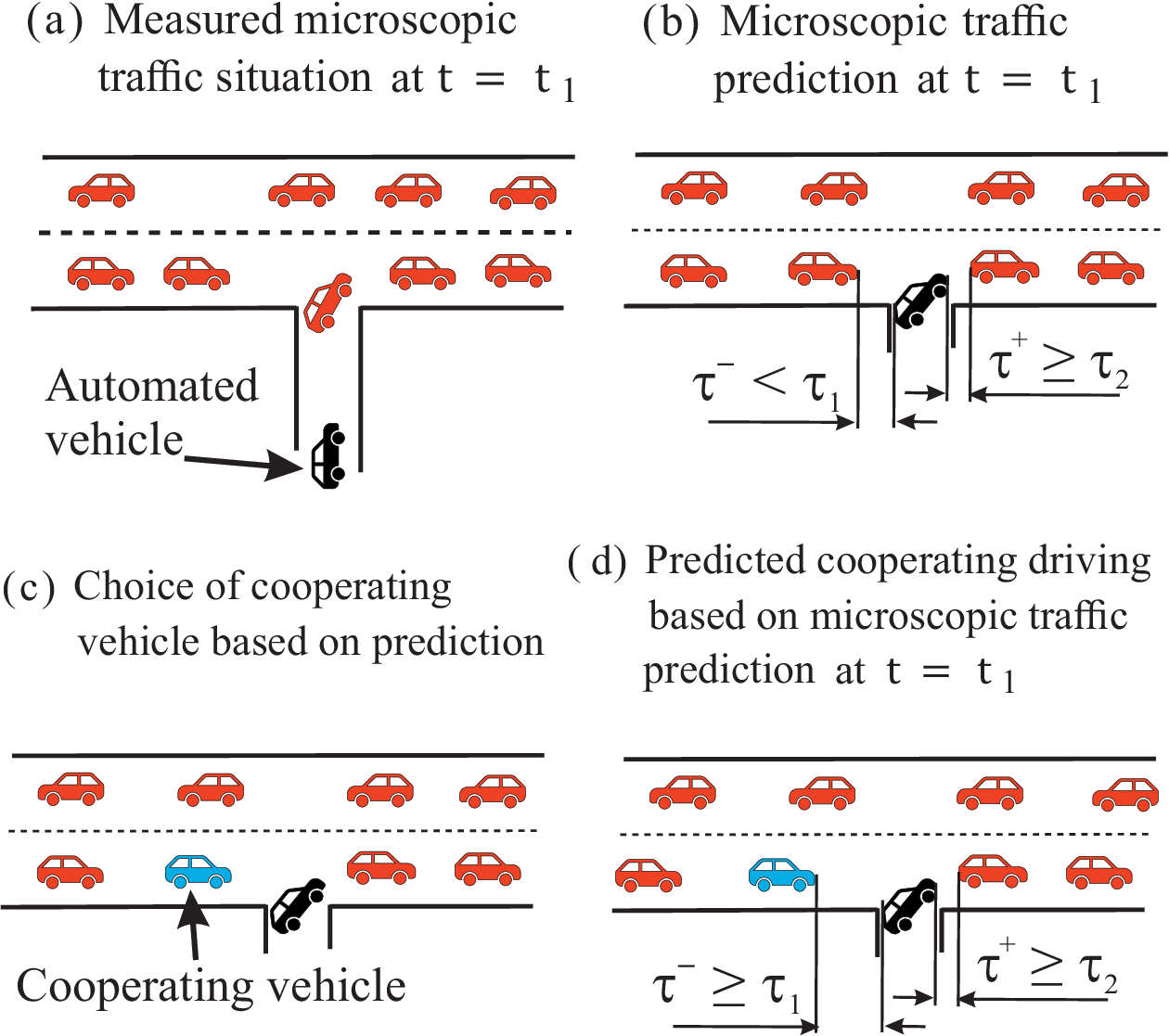}
\end{center}
\caption[]{Qualitative explanation of the methodology for cooperative driving based on microscopic traffic prediction.} 
\label{Scenario_right}
\end{figure}

In the paper, we study physical features of the above methodology  of cooperative driving through numerical simulations of a common case in city traffic in which the subject vehicle is an automated vehicle.  The automated vehicle
(black colored vehicle $\lq\lq$automated vehicle" in Fig.~\ref{Scenario_right})    moves initially on a secondary road and it would like to turn right onto a priority road. There is no traffic signal on the   intersection of these two roads. The objective of cooperative driving is to find whether there is a possibility for the merging of the automated vehicle onto the priority road {\it without stopping} at the intersection.

As mentioned, the use of cooperative driving 
 makes sense, if automated vehicle motion   control based on the microscopic traffic prediction
without cooperative driving   does not  lead to    the merging of the automated vehicle onto the priority road   without stopping  at the intersection.  Such a case is qualitatively shown in  Fig.~\ref{Scenario_right}. Indeed, when the microscopic traffic prediction is applied for automated vehicle control, we can see in Fig.~\ref{Scenario_right}(b) that
  $\tau^{-}< \tau_{1}$, i.e.,
the first of safety conditions (\ref{safety_cond_basic}) is  satisfied for  none of the pairs of vehicles moving on the priority road during   time interval $t_{\rm min}\leq t < t_{\rm max}$.
This means that   the automated vehicle must stop at the intersection.  

\begin{figure}
\begin{center}
\includegraphics[width = 8 cm]{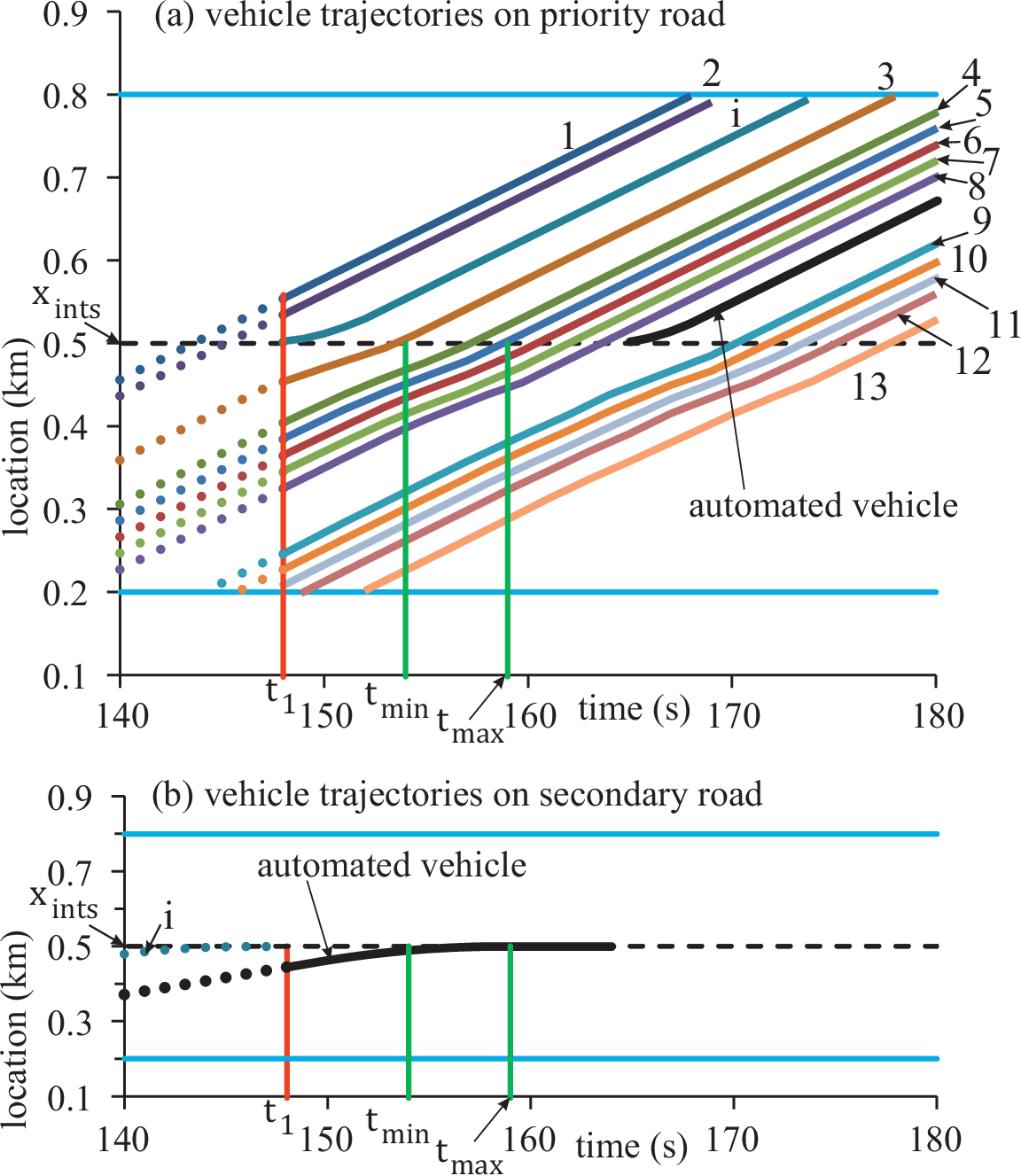}
\end{center}
\caption[]{Simulations of microscopic traffic prediction of automated vehicle motion  with approach of~\cite{KKl2022_Pred}
without the use of cooperative driving at  $t_{p}=t_{1} =$ 148 s:
  Vehicle trajectories on the priority road (a) and secondary road (b);
dotted curves are vehicle
locations of simulated microscopic traffic situations, solid curves are
trajectories of the predicted vehicle trajectories.  Parameters of the model of human driving vehicles
are given   in Table~I of   Appendix~A of~\cite{KKl2022_Pred};   
 in (\ref{safety_cond_basic}), $\tau_{1} =$ 2.0 s and $\tau_{2} =$ 0.5 s; in model of the automated vehicle
(see formulas (7)--(11) in Sec.~II of ~\cite{KKl2022_Pred}), 
we have used $\tau_{\rm d}^{\rm (AV)}=$ 1.5 s, $K_{1}=$  0.3 $\rm s^{-2}$ and $K_{2}=$ 0.6 $\rm s^{-1}$, $a^{\rm (AV)}_{\rm max}=$ 2.5 and 
$b^{\rm (AV)}_{\rm max} =$   3 ${\rm m/s^{2}}$, all other parameters of the model of the automated vehicle are the same as those in~\cite{KKl2022_Pred}. Results of calculations: $t_{\rm min}=$ 154 s and $t_{\rm max}=$ 159 s.
 A single-lane priority road is 2.5 km long; the intersection of this road with a single-lane secondary road that is 0.5 km long is at location $x_{\rm ints}$= 0.5 km; in mixed traffic flow, there are $1\%$
 of automated vehicles randomly distributed between human driving vehicles. 
Horizontal blue lines show road regions satisfying conditions (11) of~\cite{KKl2022_Pred}  within which
  microscopic traffic situations have been used for the prediction.
In simulations, the flow rate on the priority and secondary roads are, respectively, 1161 and 100 vehicles/h. 
 Poisson distribution for entering vehicles   has been used. 
}
\label{No_cooperation_t_1}
\end{figure}

 To simulate the scenario qualitatively shown in  Fig.~\ref{Scenario_right}, in
  accordance with item (i) of the general methodology of cooperative driving (Sec.~\ref{Gen}), the microscopic traffic prediction, in which no cooperative driving has been used,  is made (Fig.~\ref{No_cooperation_t_1}).
	Simulations of the predicted time instants $t_{\rm min}$ and $t_{\rm max}$,
	whose physical sense   has been explained in Sec.~\ref{Brief_MP_Sec}, is shown in Fig.~\ref{t-min_t-max}.
In the simulations, in safety conditions (\ref{safety_cond_basic}) we have used $\tau_{1} =$ 2.0 s and $\tau_{2} =$ 0.5 s.
It turns out that indeed already at the beginning of traffic prediction at $t_{p}=t_{1} =$ 148 s
condition  $\tau^{-}\geq \tau_{1}$ is  satisfied for  none of the pairs of vehicles moving on the priority road during   time interval $t_{\rm min}\leq t < t_{\rm max}$. For this reason, the automated vehicle stops at the intersection
(Fig.~\ref{No_cooperation_t_1}(b)). Only at some time instant $t > t_{\rm max}$,   safety conditions (\ref{safety_cond_basic}) are satisfied and, therefore, the automated vehicle can merge onto the priority road (Fig.~\ref{No_cooperation_t_1}(a)).

  Time instant $t_{p}=t_{1}$ of the beginning of the prediction in the simulated scenario (Fig.~\ref{No_cooperation_t_1})
			is $t_{1}=$ 148 s (vertical red line $t_{1}$ in Fig.~\ref{No_cooperation_t_1}). This is because at this time instant the preceding vehicle for the automated vehicle on the secondary road labeled by $\lq\lq$i" (Fig.~\ref{No_cooperation_t_1})
	has merged onto the priority road\footnote{In simulations, in addition to this condition,  to determine the beginning
 of the microscopic traffic prediction $t_{\rm p}=t_{\rm 1}$,  we have used the condition that the distance from the current automated vehicle location
 $x^{\rm (AV)}(t_{\rm p})$ to the location $x_{\rm ints}$ of the road intersection is less than some given value $L_{1}$
(where $L_{1}=$ 0.15 km), i.e.,
$x_{\rm ints}-x^{\rm (AV)}(t_{\rm p})<L_{1}$. This condition  is   satisfied in all simulation results presented.}. Thus, the following automated vehicle (black curves labeled by
	$\lq\lq$automated vehicle" in Fig.~\ref{No_cooperation_t_1}) can choose its deceleration freely 
	while approaching the intersection.

	\begin{figure}
\begin{center}
\includegraphics[width = 8 cm]{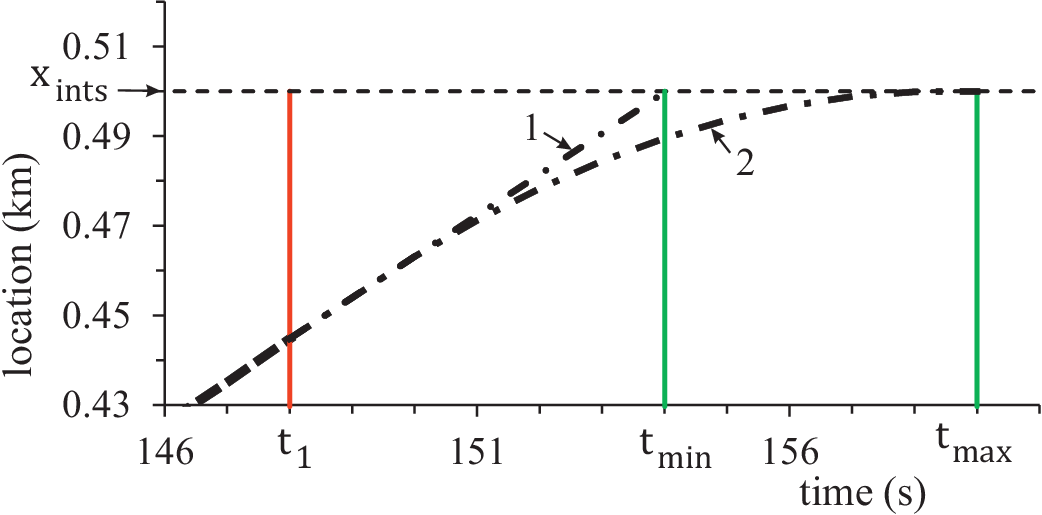}
\end{center}
\caption[]{Simulations of the  prediction of time instants $t_{\rm min}$ and $t_{\rm max}$.
Calculations are made for prediction time instant $t_{p}=t_{1} =$ 148 s:   
  Virtual trajectories of the automated vehicle on the secondary road 
	related to the prediction of $t_{\rm min}$ (dashed-dotted curve 1)
	and to the prediction of $t_{\rm max}$ (dashed-dotted curve 2), respectively.
 Other model parameters are the same as those in Fig.~\ref{No_cooperation_t_1}. 
}
\label{t-min_t-max}
\end{figure}

			The result of these calculations of microscopic traffic prediction
			made for $t_{p}=t_{1}=$ 148 s   (Fig.~\ref{No_cooperation_t_1})
	shows that in the scenario there are no pair of vehicles moving on the priority road for which {\it both}
	safety conditions (\ref{safety_cond_basic}) {\it and}
	conditions
	(\ref{t_min_t_max_t_E}) for the merging of the automated vehicle  without a stop at the intersection are satisfied.
	Therefore, the automated vehicle must stop at the intersection and wait as long as a
	pair of vehicles appears, which satisfying safety conditions
	(\ref{safety_cond_basic}), between which the automated vehicle can merge on the priority road.	
	
	Thus, in the scenario under consideration without cooperative driving
	the automated vehicle cannot
	merge on the priority road without a stop at the intersection.
 For this reason   the application of cooperative driving can have a sense.
	Simulations of the  application of cooperative driving will be considered in Secs.~\ref{Choice} and~\ref{Sim_results}.
	Before the presentation of numerical simulation results, it is useful to give
a {\it qualitative} explanation of the application of the general methodology for cooperative driving
(Sec.~\ref{Gen}) for   this case (Figs.~\ref{Scenario_right}(c, d)). First, 
in accordance of item (ii) of the general methodology,   a cooperating vehicle is chosen
(blue colored vehicle in Fig.~\ref{Scenario_right}(c)). The choice of this vehicle as the
cooperating vehicle is as follows:  From Fig.~\ref{Scenario_right}(b) we see that if this vehicle through its addition deceleration
increases time headway $\tau^{-}$ to the automated vehicle, then
 condition  $\tau^{-}\geq \tau_{1}$ can be  satisfied. In this case, the automated vehicle can merge onto the priority
road without stopping at the intersection.
Second, in accordance of item (iii) of the general methodology,
both motion requirements for the cooperating vehicle (like a required deceleration) and
related characteristics of automated vehicle control are calculated with the use of the microscopic traffic prediction
(Fig.~\ref{Scenario_right}(d)).   
Then, stages (ii) and (iii) of the general methodology explained in Figs.~\ref{Scenario_right}(c, d) for $t_{p}=t_{1}$
 are repeated at the next time instant $t_{p+1},\ p=1,2,3,\ldots$, when a next microscopic traffic situation is known, and so on. The prediction of the vehicle locations    and vehicle  speeds   are used for the calculation of the future trajectory of the automated   vehicle as well as all other vehicles involved in   cooperative driving.

 Before in Sec.~\ref{Sim_results} we consider simulation results of the scenario
 qualitatively explained in Fig.~\ref{Scenario_right}, we discuss how the cooperating vehicle is chosen in numerical simulations
 (Sec.~\ref{Choice}) as well as 
 models of motion of the automated vehicle
(Sec.~\ref{Application}) and cooperative vehicle
(Sec.~\ref{Motion_Coop})   based on microscopic traffic prediction.

	\section{Characteristics of Cooperative Driving based on Microscopic Traffic Prediction  \label{Conditions}}

	\subsection{Choice of Cooperating Vehicle based on Microscopic Traffic Prediction  \label{Choice}} 
		
At   time instant $t_{\rm 1}$, based on the calculated  microscopic traffic prediction
(Fig.~\ref{No_cooperation_t_1}), a  cooperating vehicle  is found. We assume that all vehicles can
  communicate and are willing to cooperate with the automated vehicle.  The choice of the cooperating vehicle
	is illustrated with Fig.~\ref{Choice_Fig}.

		\begin{figure}
\begin{center}
\includegraphics[width = 8 cm]{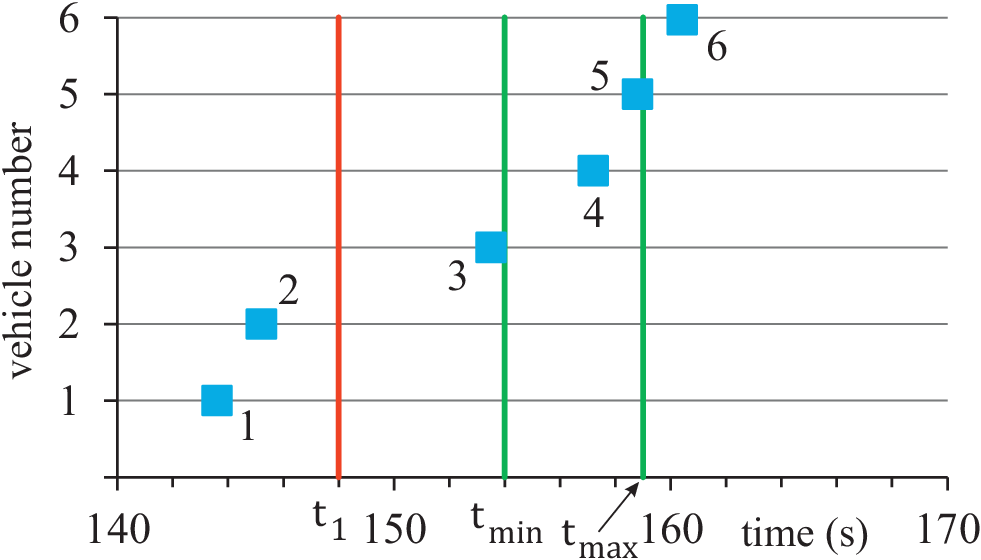}
\end{center}
\caption[]{Continuation of
Fig.~\ref{No_cooperation_t_1}. Simulations of the  prediction of time instants at which  vehicles 1--6 moving on the priority road 
reach the intersection location $x_{\rm ints}=$ 0.5 km.  Numbers 1--6 are related to vehicles 1--6
in  Fig.~\ref{No_cooperation_t_1}, respectively. 
Calculated predicted time instants at which locations of vehicles 3, 4, and 5 are
at the intersection location $x_{\rm ints}=$ 0.5 km are, respectively, $t^{(3)}=$ 153.5 s, $t^{(4)}=$ 157.2 s, and $t^{(5)}=$ 158.8 s.  
}
\label{Choice_Fig}
\end{figure}

We can see (Fig.~\ref{Choice_Fig}) that time instant, at which location of vehicle 3 is
at the intersection location $x_{\rm ints}=$ 0.5 km, is slightly smaller than
$t=t_{\rm min}$. Therefore,  vehicle 3 can be chosen as the preceding vehicle for the automated vehicle merging
onto the priority road without a stop at the intersection.
Respectively, vehicle 4 that follows vehicle 3 can be chosen as the cooperating vehicle. The
cooperating vehicle 4 should decelerate to ensure
a safety merging of the automated vehicle 
onto the priority road without a stop at the intersection. This means that
  at time instant $t=t_{\rm E}$ the cooperating  vehicle  should be behind the automated vehicle at   time gap $\tau^{-}=\tau^{(\rm c)}$ satisfying condition
\begin{equation}
\tau^{(\rm c)}\geq \tau_{1}.  
\label{t-c_t-1}
\end{equation}

\subsection{Motion of Automated  Vehicle in Cooperative Driving based on Microscopic Traffic Prediction  \label{Application}} 

 A required deceleration of the cooperating vehicle that satisfies condition (\ref{t-c_t-1}) will be considered
in Sec.~\ref{Motion_Coop}. Firstly, we should calculate  deceleration $b_{\rm p}$ of the automated vehicle
and related predicted time instant $t=t_{\rm E}$ at which the vehicle merges onto the priority road.
Predicted time instant $t=t_{\rm E}$ is found from condition  $\tau^{+}\geq \tau_{2}$ in (\ref{safety_cond_basic})
related to a safety merging of the automated vehicle behind of vehicle 3. 
The model of the calculation of the predicted time instant $t_{\rm E}$ and deceleration $b_{\rm p}$ of the automated vehicle is
the same  as that in Sec.~ III and Appendixes~B and C of~\cite{KKl2022_Pred}.

In accordance with~\cite{KKl2022_Pred}, 
  there is the repetition of the
prediction of       
the  automated vehicle trajectory   made at
each next time instants $t_{p}, \ p=2,3\ldots$. Therefore,
for each time instant $t_{p}$, the predicted time instant $t_{\rm E}$ and the associated predicted deceleration
 $b_{p}$ are calculated.  The automated vehicle motion with deceleration 
$b_{p}$ is only valid until   the next traffic prediction for $b_{p+1}$ is calculated at $t=   t_{p+1}$.
In other words,
deceleration $b_{\rm p}$ of the automated vehicle
 at time instant $t_{\rm p}$ 
 is applied only during time interval  (\ref{b_p_t_p}). 
	At $t= t_{p+1}$, the same procedure of microscopic traffic prediction as made for $t= t_{p}$
is repeated  leading to a new deceleration $b_{p+1}$
	and to a new   
  predicted  time instant $t_{\rm E}$ at which the vehicle should turn right onto the priority road.
	Then, during time interval $t_{p+1}\leq t<   t_{p+2}$ the vehicle moves in accordance with  formulas (20) and (21)
	of~\cite{KKl2022_Pred}    in which instead of the value 
$b_{p}$ the new predicted deceleration $b_{p+1}$ is used, and so on.
	Because   microscopic traffic situations  depend on time $t_{p}$,
all  predicted values, like the   predicted  time instant $t_{\rm E}(t_{p})$, are functions of  time instant $t_{p}$ at which the prediction is made.

In accordance with formula (\ref{time-interval_f}),  
time instants $t_{p}$   at which the microscopic traffic prediction is made   
can be written as
\begin{equation}
t_{p}=t_{1}+(p-1)\tau, \ p=1,2,\ldots,p_{\rm E}.
\label{t_p_max}
\end{equation}
The repetition of the
predictions of          
the  automated vehicle trajectory is made up to some maximum 
time instant $t_{p}$ denoted in (\ref{t_p_max}) by $t_{p}=t_{p_{\rm E}}$.
Conditions for the maximum  time instant $t_{p}=t_{p_{\rm E}}$ and for the related 
time instant denoted by $t^{\rm (real)}_{\rm E}$ at which the automated
 vehicle really merges onto the priority road  are presented in   
  Appendix~C   of~\cite{KKl2022_Pred}.  
 At $t\geq t_{p_{\rm E}}$
  the automated vehicle   decelerates with the last predicted deceleration
	(Appendix~C   of~\cite{KKl2022_Pred})
moving up to the road intersection at which it
   turns right  
 onto the priority road; later, no cooperating driving and no prediction is made.

\subsection{Motion of Cooperating Vehicle based on Microscopic Traffic Prediction  \label{Motion_Coop}}

\subsubsection{Deceleration of Cooperating Vehicle \label{Deceleration_cooperation}} 

For each of the time instants $t_{p}$,
the deceleration $b_{p}^{\left(\rm c\right)}$ of the cooperating vehicle that satisfies safety conditions (\ref{t-c_t-1}) is found from  condition:
\begin{equation}
x_{\rm ints}-x^{(\rm c)}_{n}=v^{(\rm c)}_{\rm E}\tau^{(\rm c)}+d+X^{(\rm c)}_{\rm b},
\label{Cooperation_condition}
\end{equation}
where   index $n$ corresponds to discrete time $n\tau,\ n=0,1,\ldots$
of the model, $x^{(\rm c)}_{n}$ is the position of the cooperating vehicle at time $t^{(\rm c)}_{n}$,
where $t^{(\rm c)}_{n}=t_{p}$; $v^{(\rm c)}_{\rm E}\geq0$ is the calculated speed of the cooperating  vehicle at time instant 
$t_{\rm E}(t_{p})$, $d$ is the vehicle length, $X^{(\rm c)}_{\rm b}$ is a deceleration distance of the cooperating  vehicle. The distance $X^{(\rm c)}_{\rm b}$ and the speed $v^{(\rm c)}_{\rm E}$  are found from the condition that the cooperating  vehicle decelerates with constant deceleration $b_{\rm p}^{\left(\rm c\right)}<0$ from the speed $v^{(\rm c)}_{n}$ to the speed $v^{(\rm c)}_{\rm E}$ during   time interval $T_{\rm c}$: 
\begin{equation}
T_{\rm c}(t_{p})=t_{\rm E}(t_{p}) - t^{(\rm c)}_{n},
\label{deceleration_time}
\end{equation} 
\begin{equation}
X^{(\rm c)}_{\rm b}=\ v^{(\rm c)}_{n}\left(T^{\rm (c)}+\delta T^{\rm (c)}\right)-
\frac{1}{2}b_{\rm p}^{\left(\rm c\right)}T^{\rm (c)}(T^{\rm (c)}+\tau+2\delta T^{\rm (c)}),
\label{deceleration_distance}
\end{equation}
\begin{equation}
v^{(\rm c)}_{\rm E}=\max\left(0,v^{(\rm c)}_{n}-b_{p}^{\left(\rm c\right)}T^{\rm (c)}\right),
\label{speed_cooperation}
\end{equation}
where $T^{\rm (c)}$ and $\delta T^{\rm (c)}$ are, respectively, the integer and the fractional parts of time $T_{\rm c}$: 
$T^{\rm (c)}=\tau\left\lfloor{T_{\rm c}}/{\tau}\right\rfloor$, $\delta T^{\rm (c)}=T_{\rm c}-T^{\rm (c)}$.
Deceleration $b_{\rm p}^{\left(\rm c\right)}$ 
is found from a solution of (\ref{Cooperation_condition})--(\ref{speed_cooperation}):  
\begin{eqnarray}
\lefteqn{b_{\rm p}^{\left(\rm c\right)}=} \nonumber\\
&{=}-\frac{2\left(x_{\rm ints}{-x}^{\rm (c)}_{n}-v^{\rm (c)}_{n}\left(T^{\rm (c)}+\delta T^{\rm (c)}
+\tau^{(\rm c)}\right)-d\right)\ }{T^{\rm (c)}\left(T^{\rm (c)}+\tau+2\delta T^{\rm (c)}+2\tau^{(\rm c)}\right)}&, 	
\label{deceleration_cooperation}
\end{eqnarray}
where 
\begin{eqnarray}
\label{b_coop}
 p=1,2,\ldots,p_{\rm E}-1. 
\end{eqnarray}
The deceleration (\ref{deceleration_cooperation}) is calculated for each time instant $t^{\rm (c)}_{n}=t_{\rm p}$. 
Then, under condition  (\ref{b_coop}) the next prediction for the time $t_{\rm E}(t_{p+1})$
(see formula (31)    of~\cite{KKl2022_Pred}) is made at time instant $t_{p+1}$, and the subsequent calculation of $b_{p+1}^{\left(\rm c\right)}$ is made.

\subsubsection{Minimum Distance for Cooperative Driving \label{min_distance}} 

At time instant $t^{\rm (c)}_{n}$ the cooperating vehicle is at some distance from the intersection location 
$D=x_{\rm ints}-x^{\rm (c)}_{n}$. It should be emphasized that 
there should be some minimum distance $D=D_{\rm min}$ for cooperative driving; in other words, cooperative driving is possible when condition 
\begin{equation}
D \geq D_{\rm min}	
\label{D_cooperation}
\end{equation}
is satisfied. 
Indeed, first the deceleration of the cooperating vehicle is limited by
  a given maximum deceleration 
$b_{p}^{\left(c\right)}=b_{\rm max}^{\left(c\right)}$. 
 In this case
$D_{\rm min}=D^{\rm (b)}_{\rm min}$, where
\begin{eqnarray}
\lefteqn{D^{\rm (b)}_{\rm min}=d+v^{\rm (c)}_{n}\left(T^{\rm (c)}+\delta T^{\rm (c)}+\tau^{\rm (c)}\right)-} \nonumber \\
&&-\frac{1}{2}b_{\rm max}^{\left(c\right)}T^{\rm (c)}\left(T^{\rm (c)}+\tau+2\delta T^{\rm (c)}+2\tau^{\rm (c)}\right).
\label{D_min_b}
\end{eqnarray}
However, $D_{\rm min}$  cannot exceed some value $D_{\rm min}^{(0)}$, because at the end of deceleration the final speed $v^{\rm (c)}_{\rm E}$ in (\ref{speed_cooperation}) cannot be negative. Therefore,   time $T^{\rm (c)}$ in (\ref{D_min_b})
 is limited by some value $T^{\rm (c)}_{\rm b}$. As a result, $D_{\rm min}$ is equal to
\begin{equation}
D_{\rm min}=\left\{
\begin{array}{ll}
D^{\rm (b)}_{\rm min} \ \textrm{at} \ T^{\rm (c)}\leq\ T^{\rm (c)}_{\rm b}
 \\
D_{\rm min}^{(0)} \ \textrm{at} \ T^{\rm (c)}>T^{\rm (c)}_{\rm b}. \\
\end{array}\right.
  \label{D_min}
  \end{equation}
In order to find   values of $D_{\rm min}^{(0)}$ and $T^{\rm (c)}_{\rm b}$ in (\ref{D_min}), a case is considered when
during some time steps $N^{\rm (c)}_{\rm b}$ the cooperating
 vehicle decelerates with the maximum deceleration $b_{\rm max}^{\left(c\right)}$ from the speed $v^{\rm (c)}_{n}$ to a
very low speed $v_{\rm c}$ that satisfies conditions $0 \leq v_{\rm c} <b_{\rm max}^{\left(c\right)}\tau$. Then,
\begin{equation}
N^{\rm (c)}_{\rm b}=\left\lfloor\frac{v^{\rm (c)}_{n}}{b_{\rm max}^{\left(c\right)}\tau}\right\rfloor,
\label{N_b}
\end{equation}
\begin{equation}
T^{\rm (c)}_{\rm b}=\tau N^{\rm (c)}_{\rm b},
\label{T_b}
\end{equation}
\begin{equation}
v_{\rm c}=v^{\rm (c)}_{n}-b_{\rm max}^{\left(c\right)}T^{\rm (c)}_{\rm b}.
\label{v_E}
\end{equation}
From  (\ref{v_E}),     value $T^{\rm (c)}_{\rm b}$ can also be written as $T^{\rm (c)}_{\rm b}={\left(v^{\rm (c)}_{n}-v_{\rm c}\right)}/{b_{\rm max}^{\left(c\right)}}$. At $T^{\rm (c)}=T^{\rm (c)}_{\rm b}$, from Eq.~(\ref{D_min_b}) we get:
\begin{eqnarray}
D_{\rm min}^{(0)}=d+\frac{1}{2}v^{\rm (c)}_{n}\left(T^{\rm (c)}_{\rm b}-\tau\right)  \nonumber  \\
 +\frac{1}{2}v_{\rm c}\left(T^{\rm (c)}_{\rm b}+\tau+2\delta T^{\rm (c)}+2\tau^{\rm (c)}\right).
\label{D_min_0}
\end{eqnarray}
Using $v^{\rm (c)}_{n}=v_{\rm free}$ (where $v_{\rm free}$ is a maximum vehicle speed)
and   condition $v_{\rm c}\ll\ v_{\rm free}$,   formulas (\ref{T_b}) and (\ref{D_min_0}) read
\begin{equation}
T^{\rm (c)}_{\rm b}=\tau\left\lfloor\frac{v_{\rm free}}{b_{\rm max}^{\left(c\right)}\tau}\right\rfloor,
\label{T_b_final}
\end{equation}
\begin{equation}
D_{\rm min}^{(0)}=d+\frac{1}{2}v_{\rm free}\left(T^{\rm (c)}_{\rm b}-\tau\right).	
\label{D_min_01}
\end{equation}
Eqs.~(\ref{T_b_final}) and (\ref{D_min_01}) determine the parameters $T^{\rm (c)}_{\rm b}$ and $D_{\rm min}^{(0)}$ in formula (\ref{D_min}) for $D_{\rm min}$. 

 Deceleration $b_{\rm p}^{\left(\rm c\right)},   \ p=1,2,\ldots,p_{\rm E}$ of the cooperating vehicle   is found   for each
microscopic  traffic prediction at time instants $t_{\rm p}$. 
The calculated deceleration $b_{\rm p}^{\left(\rm c\right)}$ can be different for different  
 time instants  $t_{\rm p},   \ p=1,2,\ldots,p_{\rm E}$.

\section{Simulation Results of Cooperative Driving \label{Sim_results}}

	At the beginning of microscopic traffic prediction $t_{1} =$ 148 s, the following predictions have been calculated: 
	\begin{itemize}
	\item [(i)]
	Time instant of automated vehicle merging $t_{\rm E}(t_{1})$ onto the priority road behind vehicle 3.
	\item [(ii)]
	The deceleration of automated vehicle $b_{\rm 1}$ (see Appendix~C    of~\cite{KKl2022_Pred}).
	\item [(iii)] A proof whether  condition $D \geq D_{\rm min}$	
(\ref{D_cooperation}) is satisfied (Sec.~\ref{min_distance}). If $\lq\lq$Yes", 
then next calculations of predicted characteristics (items (iv), (v)) are made.
	\item [(iv)] The deceleration of cooperating vehicle $b^{\rm (c)}_{\rm 1}$ (vehicle 4) (Sec.~\ref{Motion_Coop}).
\item [(v)] The speeds and trajectories of all  vehicles that locations
	correspond to conditions (11) of~\cite{KKl2022_Pred}
	  (Figs.~\ref{Coop_148s} and~\ref{AV_Veh_148s}). 
	\end{itemize}

	\begin{figure} 
\begin{center}
\includegraphics[width = 8 cm]{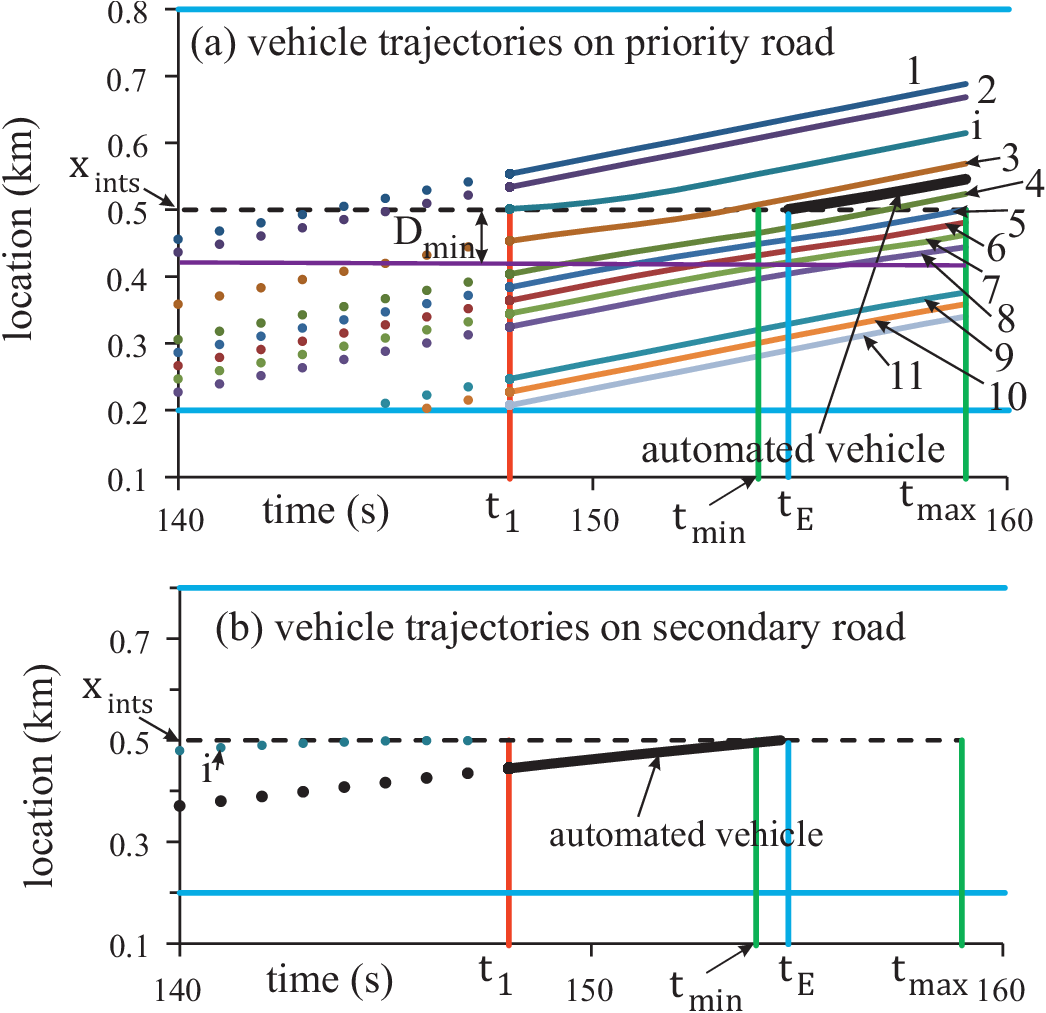}
\end{center}
\caption[]{Simulations of cooperative driving based on microscopic traffic prediction made for time instant $t_{1} =$ 148 s:
  Trajectories of vehicles  on the priority road (a) and secondary road (b). Dotted parts of vehicle trajectories are related to the microscopic traffic situation at $t\leq t_{1} =$ 148 s, solid parts of vehicle trajectories are predicted vehicle trajectories.
Model parameters: $\tau^{\rm (c)}=$  2.4 s, $b^{\rm (c)}_{\rm max}=$ 1 ${\rm m/s^{2}}$.
 Calculated values
$t_{\rm min}=$ 154  s, $t_{\rm max}=$ 159  s,   $t_{\rm E}=$ 154.6  s,   $\tau^{-}=$  2.36 s, 
 $\tau^{+}=$ 0.97  s, $D_{\rm min}=$ 78.5 m, $T_{\rm c}=$ 6.6 s,
$b_{\rm 1}=-0.21 \ {\rm m/s^{2}}$,  $b^{\rm (c)}_{\rm 1}=-0.56 \ {\rm m/s^{2}}$.
Other parameter designations,
vehicle numbers, and model parameters are the same as those in Fig.~\ref{No_cooperation_t_1}.
  }
\label{Coop_148s}  
\end{figure}

\begin{figure}
\begin{center}
\includegraphics[width = 8 cm]{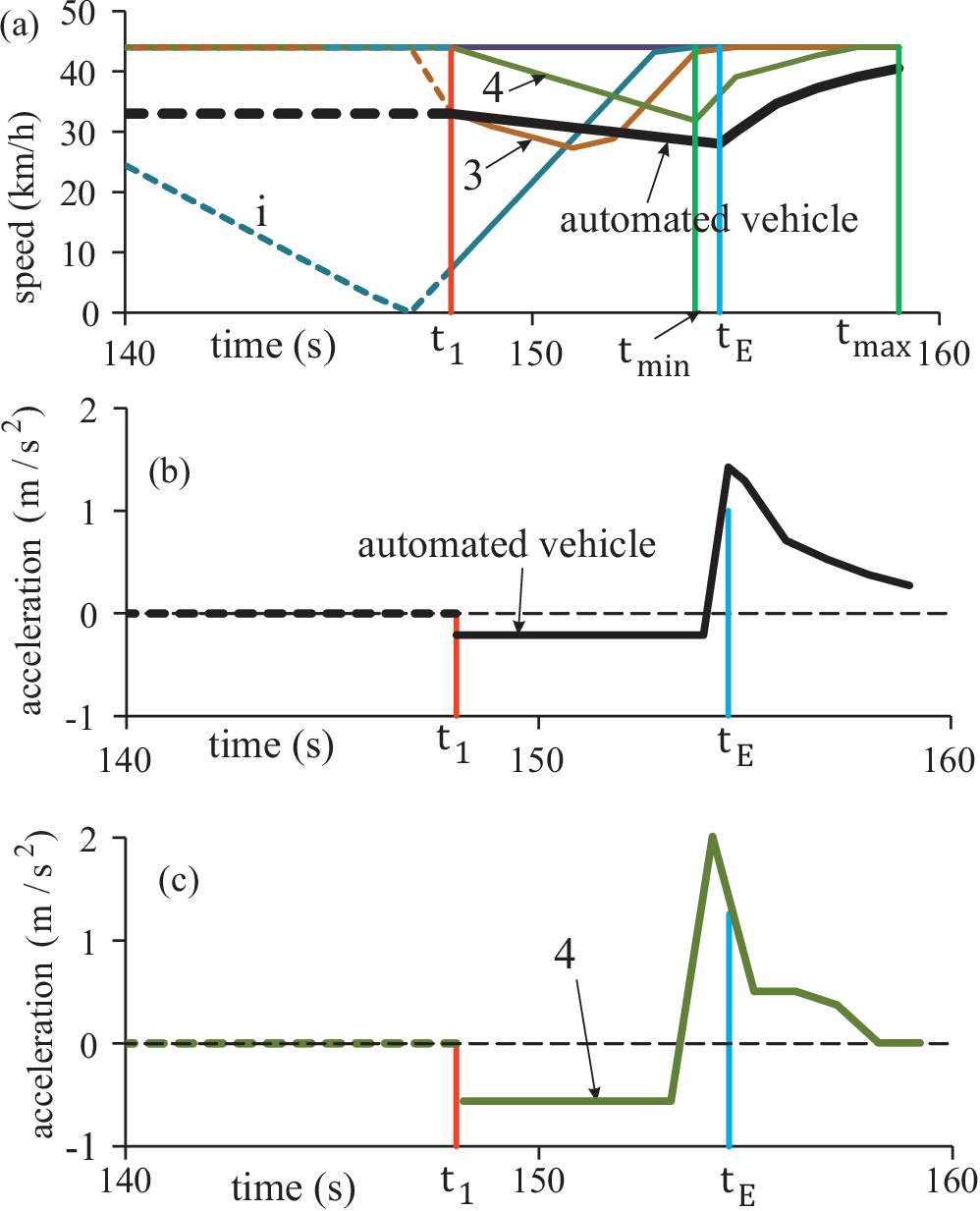}
\end{center}
\caption[]{Continuation of Fig.~\ref{Coop_148s}: (a) Time-functions of microscopic speeds
of some of the vehicles; vehicles $i$ and 3
are the preceding vehicles for the automated vehicle, respectively, on the secondary and priority roads.
(b, c) Time-functions of deceleration (acceleration) of automated vehicle (b) and the cooperating vehicle (c).
In (a, c),  vehicle 4 is the cooperating vehicle. Vehicles numbers are the same as those in Fig.~\ref{Coop_148s}.
Dashed parts of curves are related to real vehicle
motion at $t\leq t_{1} =$ 148 s, solid parts of curves are predicted characteristics.
  }
\label{AV_Veh_148s} 
\end{figure}

 Calculations of the microscopic traffic prediction at the beginning of the prediction at time instant $t_{p}=t_{1}$ 
are shown in Fig.~\ref{Coop_148s}. In this Fig.~\ref{Coop_148s},
dotted parts of vehicle trajectories are related to the microscopic traffic situation, whereas solid parts of vehicle trajectories are predicted vehicle trajectories. Time-functions of microscopic speeds
of some of the vehicles of Fig.~\ref{Coop_148s} are presented in Fig.~\ref{AV_Veh_148s}. 
 It should be emphasized that the prediction of
 values shown in Fig.~\ref{Coop_148s} and  listed in items (i)--(v) above, in particular, $t_{\rm E}(t_{1})$, $b_{\rm 1}$, and $b^{\rm (c)}_{\rm 1}$  as well as of
  trajectories of all  vehicles  have been calculated    
  during a negligible short time interval $\theta <$ 0.005 s in comparison with time step $\tau=$ 1 s of the model.   
 Corresponding to condition (\ref{b_p_t_p}), the predicted values are used by the automated vehicle and cooperating vehicle during the time interval
	\begin{equation}
t_{1}\leq t< t_{2}. 
\label{b_1_t_2}
\end{equation}

At time instant $t=t_{2}$, the next microscopic traffic situation is available. Therefore, the next 
prediction of values $t_{\rm E}(t_{2})$, $b_{\rm 2}$,  $b^{\rm (c)}_{\rm 2}$  as well as the new prediction of trajectories of all  vehicles should be calculated and used for the motion of the automated vehicle and cooperating vehicle during the next time interval
	\begin{equation}
t_{2}\leq t< t_{3},
\label{b_2_t_3}
\end{equation}
and so on.

	\begin{figure}
\begin{center}
\includegraphics[width = 8 cm]{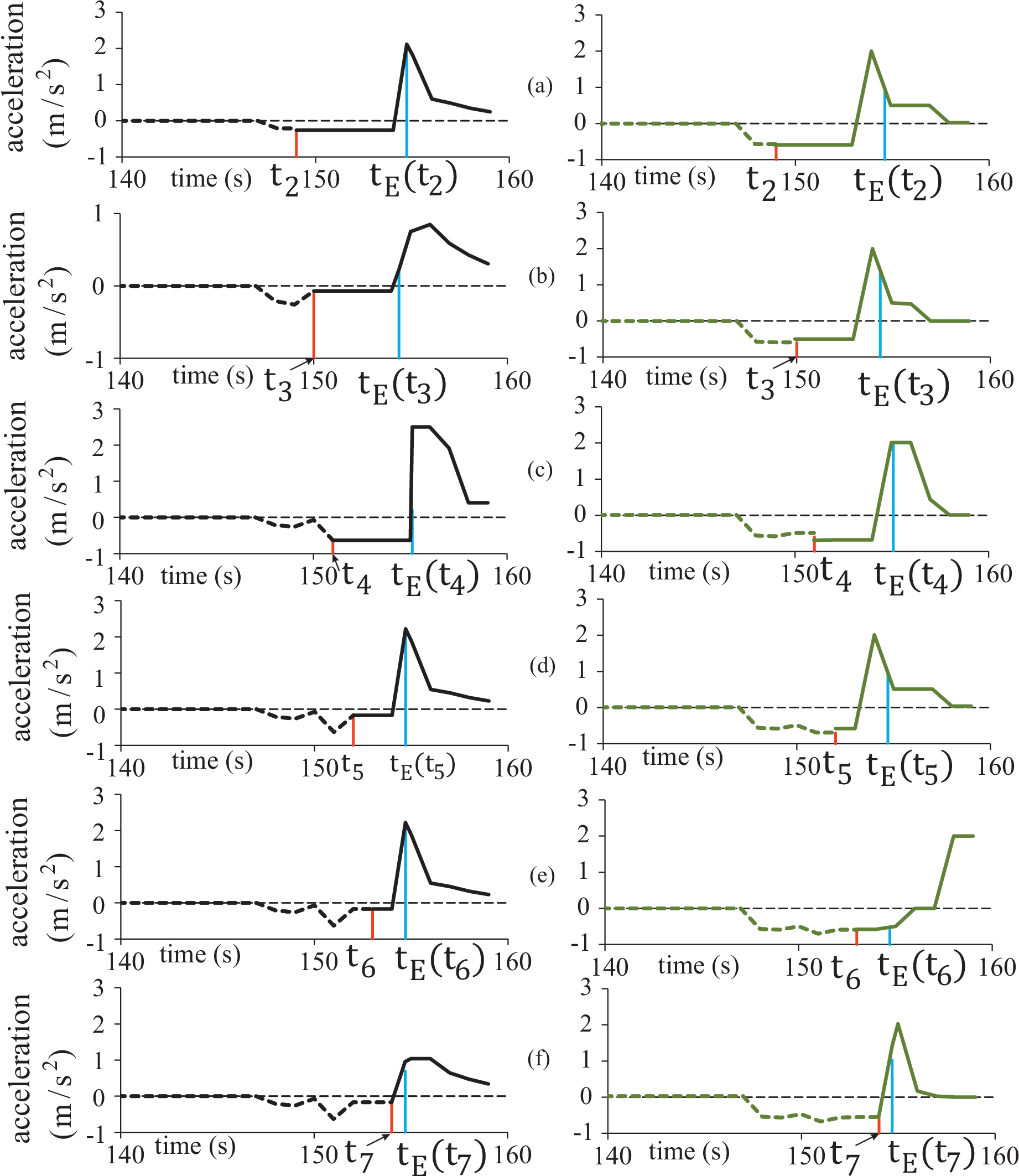}
\end{center}
\caption[]{Simulations of repetitions of prediction procedure for cooperating driving.
Time-functions of deceleration (acceleration) of automated vehicle (left column) and the cooperating vehicle (right column)
(vehicle 4  in Figs.~\ref{No_cooperation_t_1},~\ref{Coop_148s}, and~\ref{AV_Veh_148s}): (a) 
$t_{2}=$ 149 s. (b) $t_{3}=$ 150 s. (c) $t_{4}=$ 151 s. (d) $t_{5}=$ 152 s. (e) $t_{6}=$ 153 s. (f) $t_{7}=$ 154 s. Calculated predicted characteristics: $t_{\rm E}(t_{2})=$ 154.7  s, $b_{2}=-0.26 \ {\rm m/s^{2}}$, $b^{\rm (c)}_{2}=-0.59 \ {\rm m/s^{2}}$ (a),
$t_{\rm E}(t_{3})=$ 154.4  s, $b_{3}=-0.07 \ {\rm m/s^{2}}$, $b^{\rm (c)}_{3}=-0.5 \ {\rm m/s^{2}}$ (b), $t_{\rm E}(t_{4})=$ 155.1  s, $b_{4}=-0.63 \ {\rm m/s^{2}}$, $b^{\rm (c)}_{4}=-0.7  \ {\rm m/s^{2}}$ (c), $t_{\rm E}(t_{5})=$ 154.7  s, $b_{5}=-0.17 \ {\rm m/s^{2}}$, $b^{\rm (c)}_{5}=-0.59 \ {\rm m/s^{2}}$ (d), $t_{\rm E}(t_{6})=$ 154.7  s, $b_{6}=-0.17 \ {\rm m/s^{2}}$, $b^{\rm (c)}_{6}=-0.59 \ {\rm m/s^{2}}$ (e),
$t_{\rm E}(t_{7})=$ 154.7  s, $b_{7}=-0.17 \ {\rm m/s^{2}}$, $b^{\rm (c)}_{7}=-0.58 \ {\rm m/s^{2}}$ (f). Dashed parts of curves are related to real vehicle
motion at $t\leq t_{p}$, where $p= 2$ for (a), $p= 3$ for (b), $p= 4$ for (c), $p= 5$ for (d),
$p= 6$ for (e), and $p= 7$ for (f); solid parts of curves are predicted characteristics.
  }
\label{Coop_Veh_149-154s} 
\end{figure}

A dependence of predicted characteristics of the motion of the automated vehicle and cooperating vehicle
  on  time instants $t_{p}, \ p= 2,3,4,5,6,7$  
is illustrated  in  Fig.~\ref{Coop_Veh_149-154s}.  With the used of the microscopic traffic prediction, it has been found  that at
$t_{2}=$ 149 s,  to satisfy safety conditions (\ref{safety_cond_basic}),   cooperating vehicle 4 should decelerate
with $b^{\rm (c)}_{2}=-0.59 \ {\rm m/s^{2}}$; then, while decelerating with  $b_{2}=-0.26 \ {\rm m/s^{2}}$,
 the automated vehicle can safe merge onto the priority road without stopping at the intersection at
time instant  $t_{\rm E}(t_{2})=$ 154.7  s (Fig.~\ref{Coop_Veh_149-154s}(a)). 
However, both the decelerations of the cooperating vehicle 4 and  
of the automated vehicle has been applied during the time interval $t_{2}\leq t<t_{3}$   only, where $t_{3}=$ 150 s.

At $t_{3}=$ 150 s, when a new microscopic traffic situation has been measured, through the use of
 new microscopic traffic prediction  calculations of  
 the decelerations  
of the cooperating vehicle 4 and  
of the automated vehicle 
as well as the predicted time instant of the automated vehicle merging $t_{\rm E}(t_{3})=$ 154.4  s
have been made (Fig.~\ref{Coop_Veh_149-154s}(b)).  
However, both the decelerations of the cooperating vehicle 4 and  
of the automated vehicle have been applied during the time interval $t_{3}\leq t<t_{4}$ only, where $t_{4}=$ 151 s.
Calculations of the decelerations  
of the cooperating vehicle 4 and  
of the automated vehicle 
as well as the predicted time instant of the automated vehicle merging $t_{\rm E}$  have been repeated 
 for further time instants $t_{3}=$ 151 s, $t_{4}=$ 152 s, $t_{5}=$ 153 s, and $t_{6}=$ 154 s, at which
  new microscopic traffic situations have been measured, respectively (Figs.~\ref{Coop_Veh_149-154s}(c--f)). 

\begin{figure} 
\begin{center}
\includegraphics[width = 8 cm]{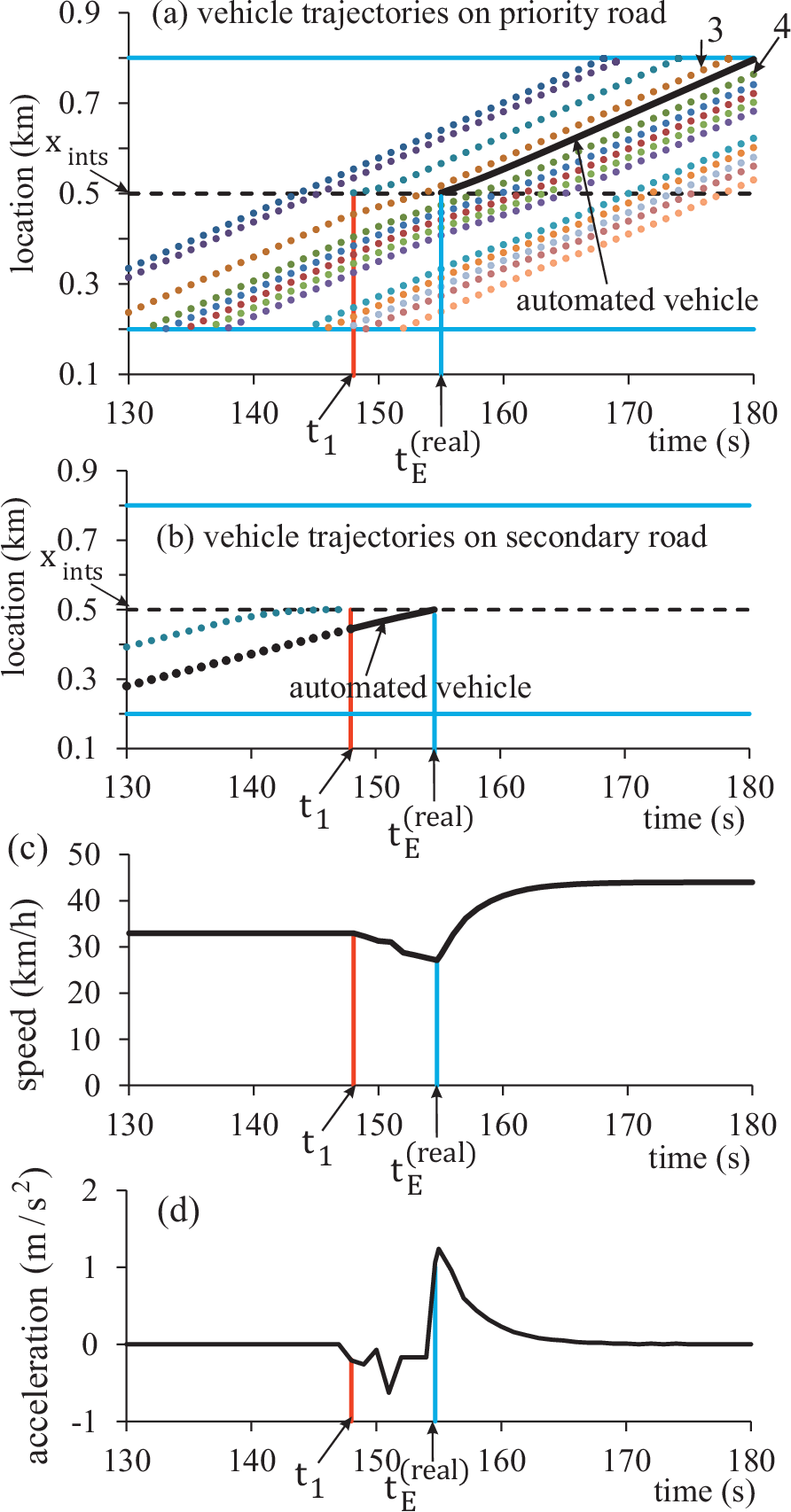}
\end{center}
\caption[]{Simulations of automated vehicle control through cooperating driving based on traffic prediction (final results):
(a, b) Vehicle trajectories on the priority (a) and secondary roads (b);
dotted   curves are related to microscopic traffic situations, solid   curves are motion of automated vehicle;
in (a), vehicles 3 and 4 are, respectively, the preceding and cooperating vehicles.
(c, d) Time-functions of speed (c) and deceleration (acceleration) of automated vehicle.
Calculated value $t^{\rm (real)}_{\rm E}=$ 154.7  s.
 Other parameter designations  and model parameters are the same as those in Fig.~\ref{No_cooperation_t_1}.
  }
\label{Final_Prog+Coop} 
\end{figure}

\begin{figure}
\begin{center}
\includegraphics[width = 8 cm]{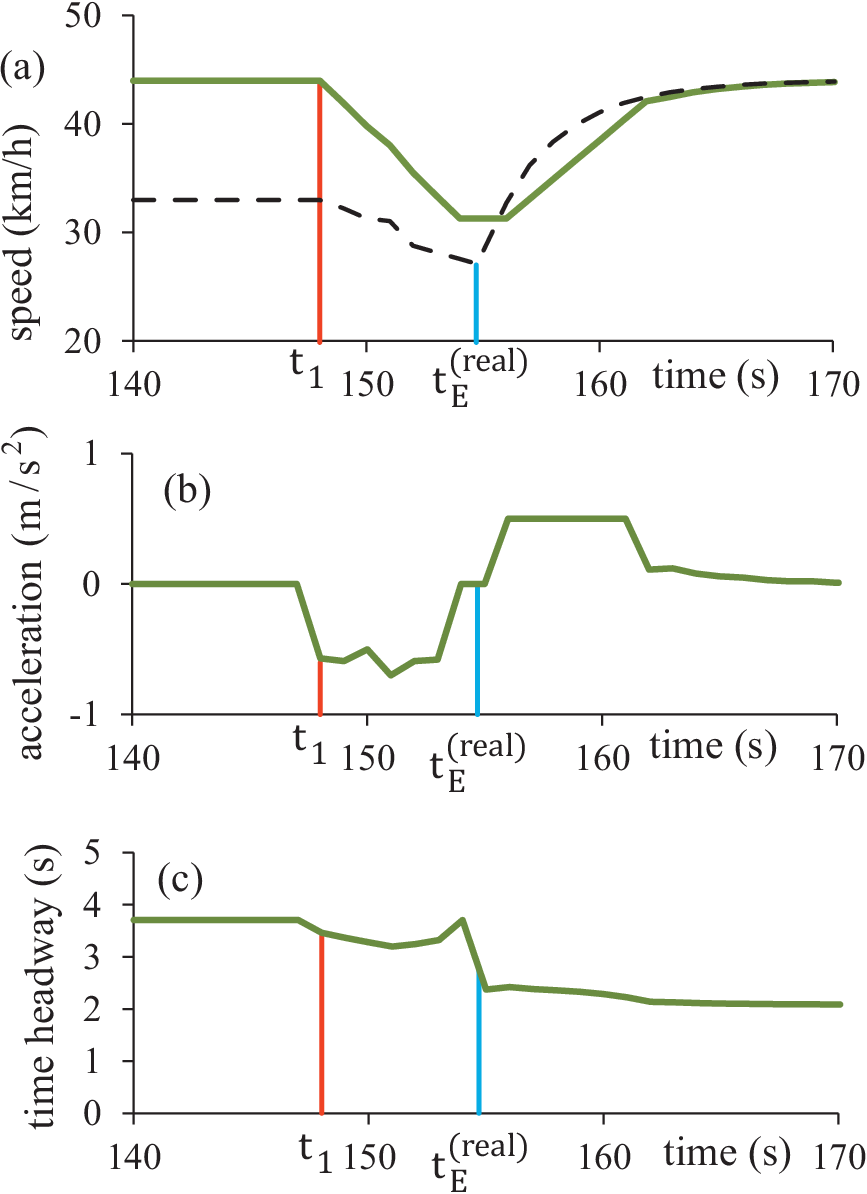}
\end{center}
\caption[]{Simulations of characteristics of motion of cooperating driving 4 based on traffic prediction (final results):
  Time-functions of speed (a),  deceleration (acceleration) (b), and time headway   of cooperating vehicle
(c). In (a), dashed curve is time-dependence of automated vehicle speed taken from
Fig.~\ref{Final_Prog+Coop}(c).
 Other parameter designations  and model parameters are the same as those in Fig.~\ref{No_cooperation_t_1}.
  }
\label{Final_Cooper_Veh} 
\end{figure}

 Thus, predicted decelerations $b_{p}$ and $b^{\rm (c)}_{p}$
as well as other predicted characteristics of vehicle motion  
can be, respectively, different  for different time instants $t_{p}, \ p= 2,3,4,5,6,7$. 
This explains why both the
automated vehicle deceleration and the deceleration of the cooperating vehicle
can be complex time functions within time interval $t_{1}\leq t<t^{\rm (real)}_{\rm E}$   
 (Figs.~\ref{AV_Veh_148s}--\ref{Final_Cooper_Veh}).

 To illustrate this conclusion, we compare
  the deceleration  
of   cooperating vehicle 4   found from microscopic traffic predictions at   time instants $t_{p}, \ p= 2,3$:
  Whereas at $t_{2}=$ 149 s cooperating
vehicle 4 decelerates with $b^{\rm (c)}_{2}=-0.59 \ {\rm m/s^{2}}$, at $t_{3}=$ 150 s
cooperating vehicle 4 should slightly reduce its deceleration to $b^{\rm (c)}_{3}=-0.5 \ {\rm m/s^{2}}$.
 The same effect is valid for the deceleration of the automated vehicle;
for example, whereas at $t_{2}=$ 149 s 
the automated vehicle should decelerate with
   $b_{2}=-0.26 \ {\rm m/s^{2}}$, however, at   $t_{3}=$ 150 s the automated vehicle should also reduce its deceleration to
$b_{3}=-0.07 \ {\rm m/s^{2}}$.

 \begin{figure}
\begin{center}
\includegraphics[width = 8 cm]{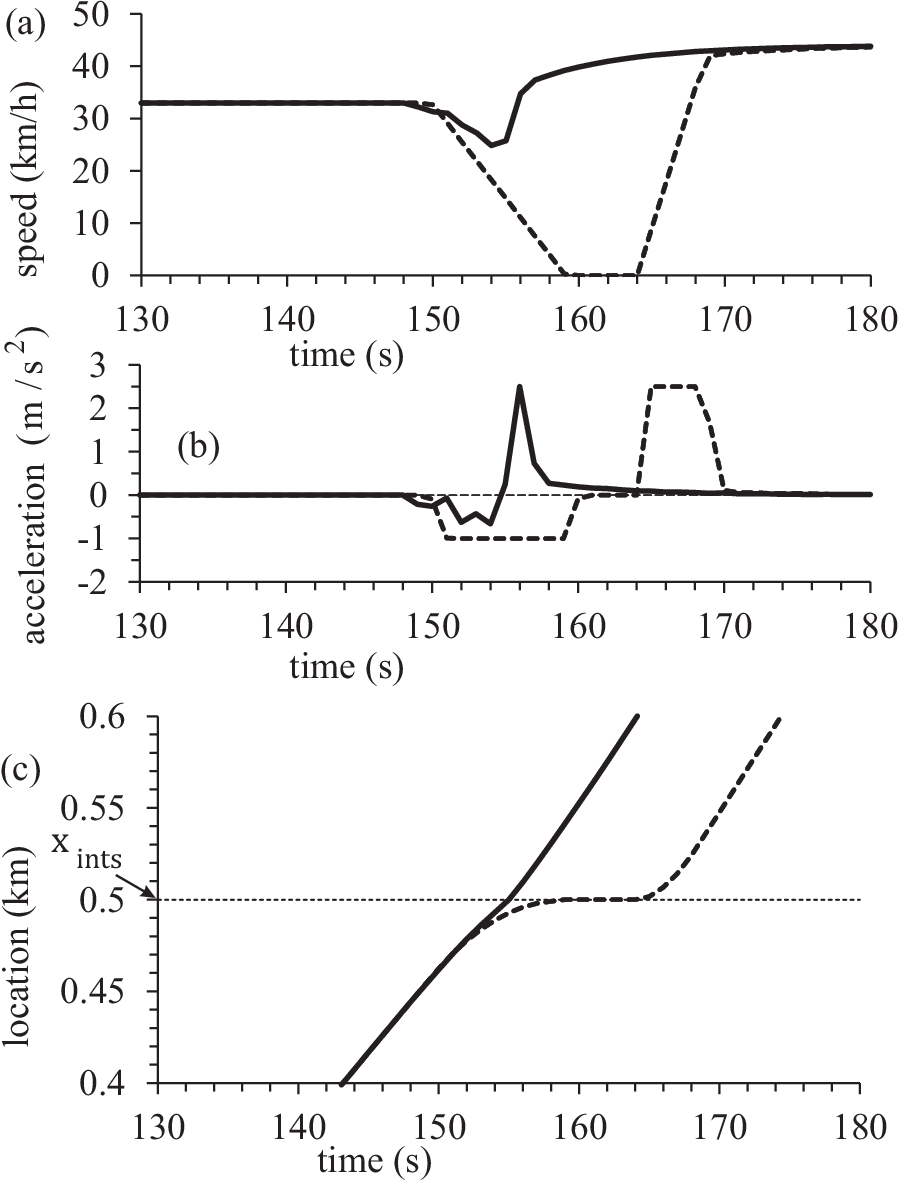}
\end{center}
\caption[]{Cooperative driving versus no cooperative driving: (a, b)
 Time-functions of speed (a) and deceleration (acceleration) (b)  of   the automated vehicle.
(c) Trajectories of the automated vehicle. Solid curves  -- automated vehicle motion
 through cooperative driving; dashed curves -- no   cooperative driving is applied.
Simulated data are taken from Fig.~\ref{No_cooperation_t_1}
for dashed curves and from Figs.~\ref{Final_Prog+Coop} and~\ref{Final_Cooper_Veh} for solid curves.
  }
\label{Coop_vs_No-Coop} 
\end{figure} 

 \begin{figure}
\begin{center}
\includegraphics[width = 8 cm]{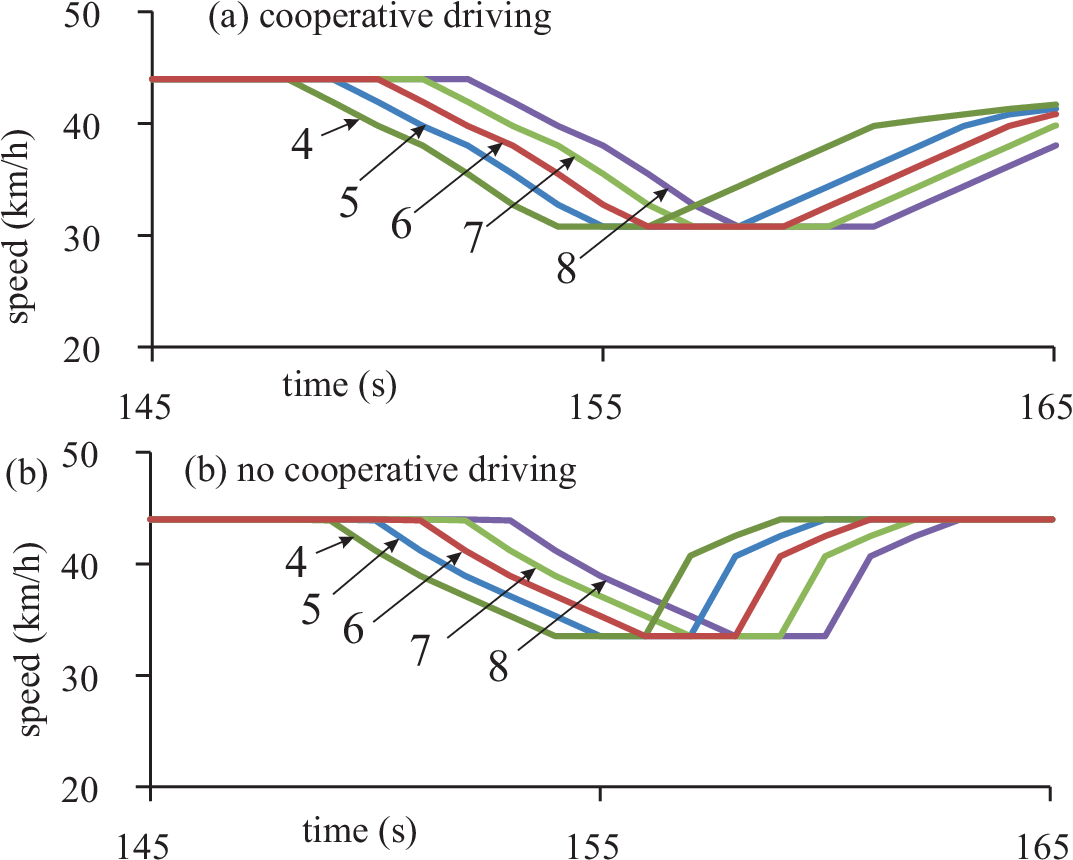}
\end{center}
\caption[]{Cooperative driving (a) versus no cooperative driving (b): Time-functions of
some of the vehicles. Simulated data are taken
from Figs.~\ref{Final_Prog+Coop} and~\ref{Final_Cooper_Veh} for (a) and
from Fig.~\ref{No_cooperation_t_1}
for  (b).  Vehicle numbers 4--8 are, respectively, the same as those in  Fig.~\ref{No_cooperation_t_1}.}
\label{Coop_vs_No-Coop2} 
\end{figure}
 
 In Figs.~\ref{Final_Prog+Coop} and~\ref{Final_Cooper_Veh}, 
simulations of cooperating driving based on {\it final}
traffic prediction are presented. As explained in Sec.~\ref{Application},   the final
traffic prediction is made at time instant $t_{p}=t_{p_{\rm E}}$
(\ref{t_p_max}). 
 At $t\geq t_{p_{\rm E}}$,
  the automated vehicle   decelerates with the last predicted deceleration
moving up to the road intersection at which it
   turns right  
 onto the priority road. Later, no cooperating driving and no prediction is made, i.e., the automated vehicle
moves on the priority road in accordance with rules of automated vehicle motion of~\cite{KKl2022_Pred}.

 Simulations of a speed harmonization 
	in city traffic through the use of cooperative driving based on the microscopic traffic prediction are shown
	in Figs.~\ref{Coop_vs_No-Coop} and~\ref{Coop_vs_No-Coop2}. 
As can be seen from Fig.~\ref{Coop_vs_No-Coop},
  the use of cooperative driving based on the microscopic traffic prediction leads to a speed harmonization 
	in city traffic.  However,
	  the deceleration of the cooperating vehicle (vehicle 4) forces the following vehicles 5--8 also to decelerate
		(Fig.~\ref{Coop_vs_No-Coop2}).
		Nevertheless, without cooperative driving   the minimum speed of vehicles 4--8 
		is 33.5 km/h (Fig.~\ref{Coop_vs_No-Coop2}(b)), whereas due to cooperative driving this minimum speed
		decreases only to 31.3 km/h (Fig.~\ref{Coop_vs_No-Coop2}(a)). Thus, the decrease in the speed in vehicles following the automated vehicle caused by cooperative driving can be very low (in the case under consideration it is only 2.2 km/h) in comparison with
		a considerably larger effect of cooperative driving on the motion of the automated vehicle (Fig.~\ref{Coop_vs_No-Coop}).

\section{Discussion \label{Disc}}

\subsection{About Limitations of  Cooperative Driving \label{Sec_Limitation}}  

	As  shown above,  the use of cooperative driving based on the microscopic traffic prediction leads to a speed harmonization 
	in city traffic.  However, there is a considerable limitation of the applicability of cooperative driving.
		
\begin{figure}
\begin{center}
\includegraphics[width = 8 cm]{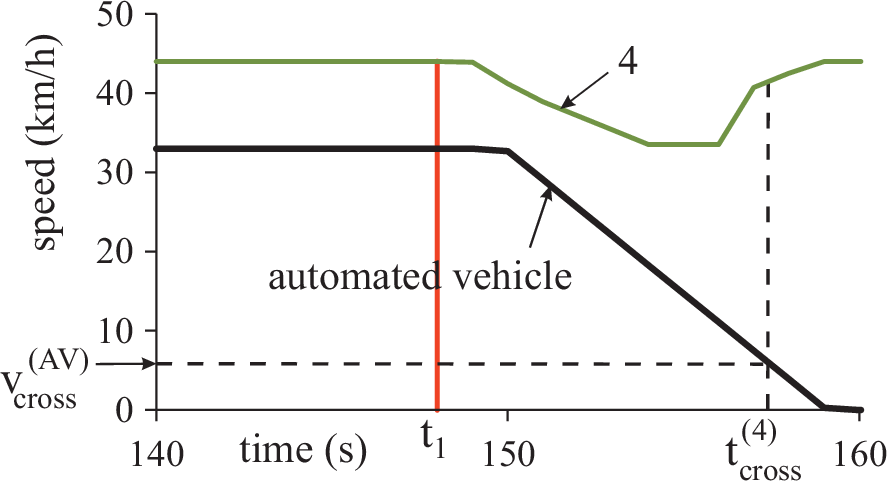}
\end{center}
\caption[]{Explanation of limitations of  cooperative driving.
 Time-functions of speed   of vehicle 4  and the automated vehicle along trajectories of these vehicles
presented in Fig.~\ref{No_cooperation_t_1}. Calculated values: $t^{\rm (4)}_{\rm cross}=$ 157.2  s,
$v^{\rm (AV)}_{\rm cross}=$ 3.89  km/h.
  }
\label{Veh_5} 
\end{figure}
 
To understand this limitation, we assume that vehicle 4 in Fig.~\ref{No_cooperation_t_1}
cannot cooperate or it does not want to cooperate.
We assume that the next upstream vehicle 5 can  cooperate and it    wants to cooperate.
We would like to understand in which degree the cooperation of vehicle 5 can improve 
traffic flow characteristics. 

With this objective, we consider Figs.~\ref{Choice_Fig} and~\ref{Veh_5}.
Because we have assumed that 
vehicle 4 does not cooperate, the automated vehicle should merge onto the priority road behind vehicle 4.
Therefore, the automated vehicle must decelerate before vehicle 4  
 is located already downstream of the intersection to satisfy safety condition 
$\tau^{+}\geq \tau_{2}$.
However, 
from  Fig.~\ref{Choice_Fig} we can see that vehicle 4 reaches the intersection location $x_{\rm ints}=$ 0.5 km at
time instant $t^{\rm (4)}_{\rm cross}=$ 157.2 s (Fig.~\ref{Veh_5}).
At this time instant the automated vehicle speed is already equal to a very low value 
$v^{\rm (AV)}_{\rm cross}=$ 3.89 km/h (Fig.~\ref{Veh_5}).
Thus, the automated vehicle can merge onto the priority road at some time instant $t_{\rm E}>t^{\rm (4)}_{\rm cross}$
while decelerating further to some lower speed $v^{\rm (AV)}_{\rm E}<$ 3.89 km/h that is close to zero.
Although even in this case of cooperative driving
the automated vehicle can merge onto the main road without a stop at the intersection,
 it is realized on the cost of the deceleration of vehicle 5 and subsequent deceleration of
some other following vehicles 
on the priority road.
It seems that such cooperative driving is not effective enough to be used in the reality.

We see that even if in traffic flow only a few of the vehicles cannot  cooperate or they do not   want to cooperate,
this can lead to considerable decrease in the effective   use of cooperative driving.
For this reason, a statistical analysis of cooperative driving based on microscopic traffic prediction for the case when some of the vehicles do not cooperate
 could be a very interesting task for further traffic studies.
However, such a statistical analysis is a separate study, which is out of the scope of this paper.
As above-mentioned (Sec.~\ref{Int}), the objective of the paper is   limited to the presented   methodology  of cooperative driving based on microscopic traffic prediction in mixed traffic flow
with 100$\%$ of vehicles that  participate in  cooperative driving.

\subsection{Short-Time  Prediction:
Statistical Algorithms of Artificial Intelligence (AI) versus Microscopic Physical  Traffic Modeling  \label{MatMod_vs_AI}}

 Leaning algorithms of AI are   the   important basis
of many future technologies. This is also the case for future transportation technologies.
In particular, as mentioned,  in traffic science the short-time prediction of  vehicle variables (vehicle locations
and speeds)   in the neighborhood of the automated vehicle, which is required for the planning of the trajectory of the automated  vehicle, is made based on a diverse variety of  statistical approaches of AI. Some of these statistical approaches  have been mentioned above~\cite{Brechtel2014,Lin2019,Schorner2019,Klimenko2014,Hubmann2019,A-Lombard-A_2023A,J-Luo-T_2023A,Markov_Shi_2024A,Isele2018,Qiao2018,Sama2020,D-Chen_2023A,X-Jiang-J_2023A,Learn_Pred_AV_2024A,Zhou-Z-Cao_2023A,Zhang-S-Li_2023A,Shupei-Wang_2024A,Renteng-Yuan_2023A,Du-Y-Zou_2023A,K-Guo-M-Wu_2023A,Chunyu-Liu_2023A,J-Guo_2023A,Bharti-Redhu_2023A,Wensong-Zhang_2023A,Shun-Wang_2023A,Siyuan-Feng_2023A,Li-Z-Huang_2023A,Jiaxin-Liu_2023A,Zelin-Wang_2024A,C-J-Hoel_2023A,S-Fang-C_2023A,Orzechowski2018,Althoff2016,Naumann2019,Tas2018,Akagi2015,Morales2017,Yoshihara2017,Takeuchi2015,Yu2019,Hoermann2017,Guojing-Hu_2023A,Changxi-Ma_2023A,S-Akhtar_2023A,Xiaoxue-Yang_2023A,Gao-X-Li_2023A,Z-Gu_2023A,Y-Zhang_2023A,Xiaoyong-Sun_2023A,Weibin-Zhang_2023A,H-Chen-Y-Liu_2023A,W-Shao_2023A,T-Qie-W_2023A}. 

 In the paper, rather than one of the  statistical approaches of AI,  
we follow a qualitative different approach to traffic prediction~\cite{KKl2022_Pred} in which   no statistical analysis of a historical  traffic database   is used for traffic prediction. To explain our motivation, we should  explain some basic physical traffic features.
First, traffic occur in time and space. Second, vehicle locations and speeds  are continuous time-variables.
Third,   traffic prediction needed for the confident  planning of the trajectory of the automated  vehicle  
 can include a large enough number of vehicles whose locations and speeds   should be predicted. 
Therefore, we can assume the following
  basic problems for   applications of the  statistical approaches of AI for  the planning of the trajectory of the automated  vehicle in real traffic:
\begin{enumerate}
\item [(i)] For each part of a traffic network
and even at the same initial traffic conditions at the network part boundaries, there can be {\it infinity number of   microscopic
traffic situations}.  Therefore, a statistical database
cannot include all possible  microscopic
traffic situations. This is true independent of the time duration within which traffic data included in
this database have been measured.
\item [(ii)] Because the statistical database
cannot include all possible  microscopic
traffic situations (item (i)), in general case, the probability of the prediction of a particular
 microscopic traffic situation with a given high enough accuracy needed for {\it  the safety planning} of automated vehicle trajectory is probably very difficult to  determine.
	\item [(iii)] Due to differences in road infrastructure of different network parts within which
	the short-time prediction of  vehicle variables     in the neighborhood of the automated vehicle is made,
	it can occur that different historical traffic databases are needed for each of the network parts~\footnote{Examples of the differences in road infrastructure include different numbers of on- and off-ramps and distances between them, road gradients, road curves, changes in the number of road lanes, etc.}.
\end{enumerate} 

  All these
problems  of the  statistical approaches of AI for  short-time microscopic traffic prediction in real traffic are absent in the approach to traffic prediction of~\cite{KKl2022_Pred}. This is because in this approach rather than statistical analysis of a historical  traffic database, the microscopic modeling of the
 motion of each of the vehicles is used for microscopic traffic prediction. We see the following advantages of cooperative driving based on microscopic traffic prediction:
\begin{enumerate}
\item [1.] No statistical database for microscopic traffic prediction is needed.  
\item [2.] A microscopic traffic flow model based on measured microscopic traffic situations (locations and speeds of vehicles) can predict future microscopic traffic situation.
In accordance with this prediction, a decision for the motion of 
the cooperating vehicles and  the safety planning of automated vehicle trajectory can be made.
	\item [3.] Parameters of the   microscopic traffic flow model can include
	a variety of vehicle behaviors adjustable for  different  road infrastructure, like on- and off-ramps, road curves and road gradients etc. Therefore, in general there is no need to use
	different traffic models for each of the network parts.
\end{enumerate}

 It must be emphasize that  statistical approaches of AI can also be very useful for the further development of the
microscopic traffic prediction approach of~\cite{KKl2022_Pred}. Indeed,
parameters of a microscopic traffic model used for the  prediction
can be automatically adapted over time with the use of
learning approaches of AI.
However, this task, which can be very interesting for further traffic studies,  is out of the scope of this paper.

\subsection{Other Applications of Methodology for Cooperative Driving  \label{Other_Sec}}

 We have  illustrated and studied numerically the  methodology of cooperative driving
for a particular simple traffic scenario, in which a subject vehicle will to turn right at a unsignalized city intersection (Sec.~\ref{Sec_Scenario}). From the stages of the methodology (Sec.~\ref{Gen}), it is clear that its applicability 
for   other traffic scenarios is limited by the applicability of the microscopic traffic prediction
of~\cite{KKl2022_Pred}. In a general case, the prediction of the motion of the local
neighbors for the subject vehicle can depend on the
behavior of the subject vehicle, whereas the behavior of
the subject vehicle depends on the motion of the neighbor
vehicles. 
 In other words, without cooperative driving
the microscopic traffic prediction 
 can be applied  at least when  the predicted motion of the
local neighboring vehicles does not depend on the subject
vehicle motion.  As explained in Sec.~VII A of~\cite{KKl2022_Pred}, there can be a number of other other
possible applications of the microscopic traffic prediction  in which these conditions are satisfied. Based on the prediction, cooperative vehicles and their behavioral parameters  are chosen as explained in Sec.~\ref{Gen}. 

	\begin{figure}
\begin{center}
\includegraphics[width = 8 cm]{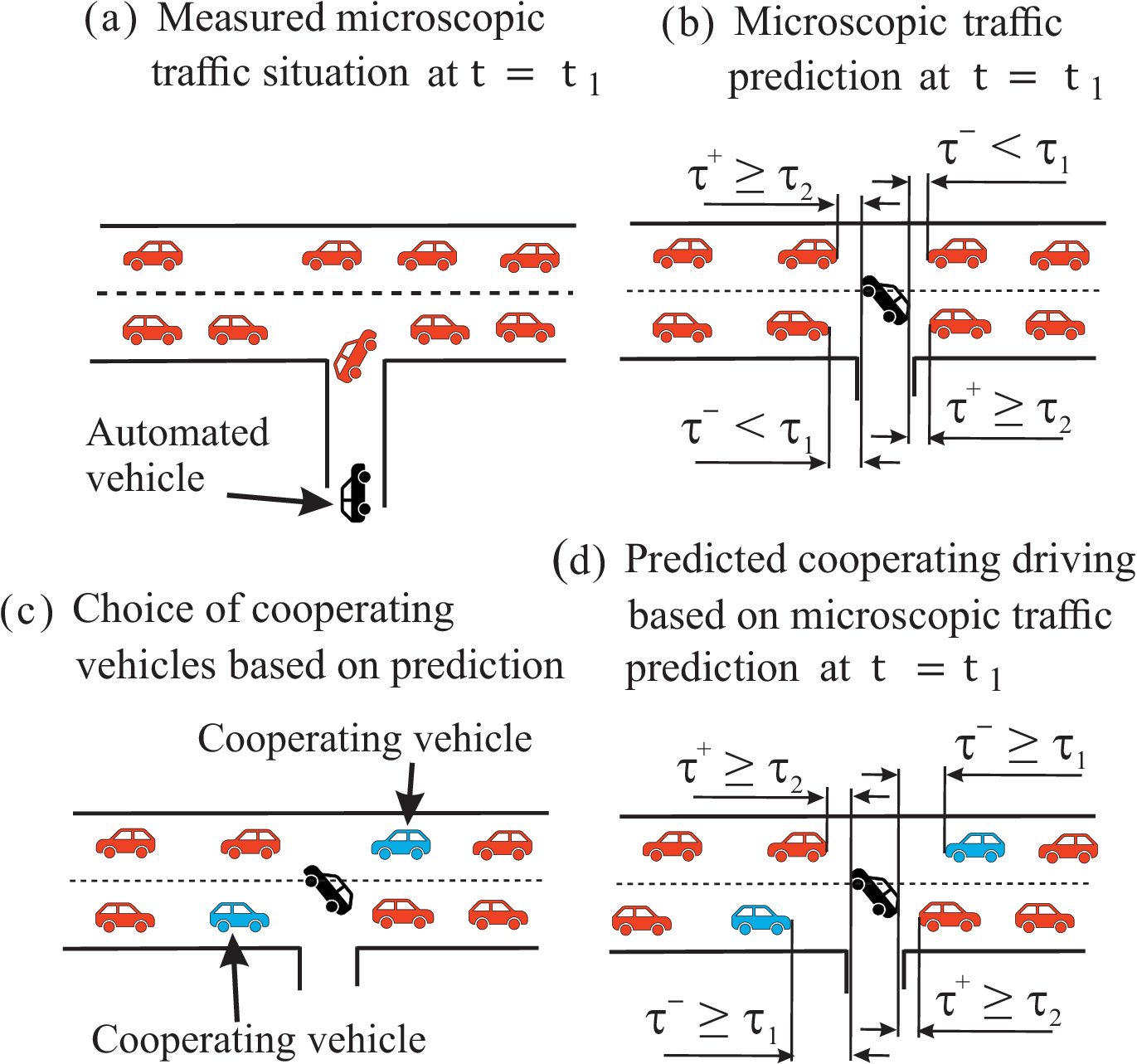}
\end{center}
\caption[]{Qualitative explanation of the application of the methodology for cooperative driving based on microscopic traffic prediction for a traffic scenario, in which a subject vehicle will to turn {\it left} at the unsignalized city intersection.  }
\label{Scenario_left}
\end{figure}

 As follows from Sec.~\ref{Gen}, the physical basis of 
   the   methodology for cooperative driving presented in the paper is 
	the microscopic traffic prediction approach of~\cite{KKl2022_Pred}. Indeed, the prediction approach
	  is used for all states of the application of cooperative driving that are as follows:

	(i) To make a decision whether cooperative driving is required for vehicle trajectory planning.

	(ii) To make a  choice of required
	cooperative vehicles.

	(iii) To calculate the motion behaviors of the cooperative vehicles that satisfy 
	driving safety conditions.

The microscopic traffic prediction approach of~\cite{KKl2022_Pred} used in the methodology of cooperative driving exhibits 
 a general character. For this reason, the methodology of cooperative driving presented in this paper
	can be applied for different traffic scenarios in which 
	the prediction approach is applicable. To explain these conclusions, below we consider qualitatively
	one of the  traffic scenarios (Fig.~\ref{Scenario_left}).

  Contrary to the simple
 scenario studied above, in which the automated vehicle tries to turn {\it right} at the unsignilized intersection
(Fig.~\ref{Scenario_right}), in the new considerably more complex scenario the automated vehicle
tries to turn {\it left} at this intersection
(Fig.~\ref{Scenario_left}).

 In accordance with item (i) of the general methodology (Sec.~\ref{Gen}), we assume that in a traffic situation of this scenario
(Fig.~\ref{Scenario_left}(a)) the application of microscopic traffic prediction
without cooperative driving   does not  lead to    the turning left of the automated vehicle onto the priority road   without stopping  at the intersection. This case is qualitatively shown in Fig.~\ref{Scenario_left}(b), in which
the first of safety conditions (\ref{safety_cond_basic}) is not satisfied: Condition
$\tau^{-}< \tau_{1}$ is valid for both right and left traffic directions.

 In accordance of item (ii) of the general methodology,     cooperating vehicles are chosen
(blue colored vehicles in Fig.~\ref{Scenario_left}(c)). The choice of these cooperating vehicles   is as follows:  From Fig.~\ref{Scenario_left}(b) we see that if these vehicles through their addition deceleration
increase time headway $\tau^{-}$ to the automated vehicle, respectively, in the right and left directions, then
 condition  $\tau^{-}\geq \tau_{1}$ can be  satisfied for both directions. Therefore, the automated vehicle can merge onto the priority
road without stopping at the intersection.

In accordance of item (iii) of the general methodology,
 motion requirements for the cooperating vehicles (like   required decelerations) and
related characteristics of automated vehicle control are calculated with the use of the microscopic traffic prediction
(Fig.~\ref{Scenario_left}(d)).   
Then, stages (ii) and (iii) of the general methodology explained in Figs.~\ref{Scenario_left}(c, d) for $t_{p}=t_{1}$
 are repeated at the next time instant $t_{p+1},\ p=1,2,3,\ldots$, when a next microscopic traffic situation is known, and so on. The prediction of the vehicle locations    and vehicle  speeds   are used for the calculation of the future trajectory of the automated   vehicle as well as all other vehicles involved in   cooperative driving.

These qualitative conclusions should not be necessarily to be proven through   numerical simulations of the traffic scenario 
shown in Fig.~\ref{Scenario_left}. This is due to the general character of the microscopic
traffic prediction approach of~\cite{KKl2022_Pred} used for the methodology of cooperative driving
of this paper. Indeed,   the procedure of the   prediction of values
		$t=t_{\rm min}$,  $t=t_{\rm max}$, and $t=t_{\rm E}$ in (\ref{t_min_t_max_t_E})
		for both potential pairs of vehicles moving in the right and left traffic directions on the priority road
		as well as the procedure of the calculation of requirements for the cooperating vehicles  and
  automated vehicle control
		presented in~\cite{KKl2022_Pred} are general ones. This means that   the microscopic prediction 
		is independent of whether right or left traffic direction is considered.
		The   difference of the scenario $\lq\lq$left turning" (Fig.~\ref{Scenario_left}(b)) in comparison with
		the scenario $\lq\lq$right turning" (Fig.~\ref{Scenario_right}(b)) is  as follows: In Fig.~\ref{Scenario_left},
		safety conditions (\ref{safety_cond_basic}) should be proven    for {\it both} the right
		{\it and} left traffic directions\footnote{It can occurs
		that rather than the first of safety conditions (\ref{safety_cond_basic}) is not satisfied for both right and left traffic directions (Fig.~\ref{Scenario_left}(b)), this safety condition is not satisfied only for one of the directions.
		Then, only one cooperating vehicle should be chosen for the traffic direction for which the first of safety conditions (\ref{safety_cond_basic}) is not satisfied.}. 
 
 Thus, due to a general character of the microscopic traffic prediction approach of~\cite{KKl2022_Pred},
   the   methodology presented in the paper can be used for different traffic scenarios in which 
	the prediction approach is applicable. For this reason,   the physics of the   methodology
	for cooperative driving could be understood from simulations of a simple scenario of right turning (Fig.~\ref{Scenario_right})
	as made in the paper.
  Further simulations of the general methodology of cooperative driving for more complex scenarios can be very interesting for traffic engineering that are out of the scope of the paper.

\subsection{Effect of Data Uncertainty on Prediction Reliability  \label{real_Sec}}

 Because measurements of microscopic traffic situations for
unsignalized intersections are not available, microscopic traffic situations have been simulated with the same microscopic traffic flow model shortly discussed in Sec.~\ref{Model_subsec}. 
Moreover,
we have assumed that vehicle speeds and locations
in microscopic traffic situations that are used as initial
conditions for the microscopic traffic prediction at time instants $t_{p}, \ p=1,2,\ldots $
(see formula (22) of~\cite{KKl2022_Pred}) have no errors in comparison with real vehicle
speeds and locations of microscopic traffic situations.

As shown in~\cite{KernerBook,KernerBook2},  the Kerner-Klenov microscopic stochastic traffic flow model used in all simulations in this paper 
is able to simulate both microscopic and macroscopic measured (empirical) traffic data with
a sufficient accuracy\footnote{See also references to origin papers
 cited
in the books~\cite{KernerBook,KernerBook2} about the
empirical proof of the Kerner-Klenov microscopic stochastic traffic flow model.}. Nevertheless, the following   
  question can arise: Is  
there    a potential discrepancy between the predicted vehicle trajectories and the actual trajectories.
Indeed, in reality there can be data latency errors and/or random
errors in measurements of the vehicle speeds and locations in
microscopic traffic situations used in the microscopic traffic
prediction. We can call the errors the uncertainty of traffic
data used in the prediction model.  	 

 The related question how the
 robustness of the prediction model is maintained in the face of significant prediction errors caused by 
the uncertainty of traffic
data  has already been addressed in details in Secs.~IV--VIII of~\cite{KKl2022_Pred}. 
 It has been found that the microscopic traffic prediction can guarantee collision
avoidance and safety traffic even when there is a
considerable data uncertainty caused by data latency, random
errors of the vehicle locations, and/or random errors of the
vehicle speeds in the data of microscopic traffic situations. However, there is 
  a critical uncertainty, i.e., the maximum
amplitude of errors in the data of microscopic traffic situations
at which the microscopic traffic prediction can still reliably be
used for the automated vehicle control. There is probability $P_{\rm app}$ of the reliability of the application of
the microscopic traffic prediction:
when the data uncertainty does not exceed the critical
uncertainty, then probability $P_{\rm app}=$ 1~\cite{KKl2022_Pred}. In this case, the microscopic
traffic prediction is applicable for automated vehicle
control.  

\subsection{Conclusions \label{Conc}} 

  1. We have introduced a methodology for cooperative driving in vehicular traffic
based on time-limited  microscopic traffic prediction that is performed based on the microscopic traffic modeling.  
Contrary to most approaches based on statistical algorithms of AI, no historical database is required for our approach. 

  2. With the use of microscopic traffic simulations of a simple city traffic scenario, we have proven the availability of the introduced
 methodology for the cooperative driving and studied its physical features.

3.  Cooperative driving based on microscopic traffic prediction
  enables the control of the automated driving vehicle in mixed traffic  in  a complex city traffic scenario
	leading to a speed harmonization and, therefore, to increase in traffic comfort and safety 
	in city traffic.
	 
	4. Due to a general character of the microscopic traffic prediction approach of~\cite{KKl2022_Pred},
   the   methodology presented in the paper can be used for different traffic scenarios in which 
	the prediction approach is applicable. 

5.  If the rate of communicating and cooperative vehicles in traffic flow becomes significantly less than 100{\%}, the possibility of cooperative driving is likely to be very limited.

 \begin{acknowledgments} 

We thank our partners for their support in the project $\lq\lq$LUKAS -- Lokales Umfeldmodell f{\"u}r das Kooperative, Automatisierte Fahren in komplexen Verkehrssituationen" funded by the German Federal Ministry for Economic Affairs and Climate Action.

\end{acknowledgments}

\end{document}